\documentclass[apj]{emulateapj}

\newcommand {\Lya}    {Ly$\alpha$}   
\newcommand {\Lyb}    {Ly$\beta$}      

\newcommand {\HI}     {\ion{H}{1}}   
\newcommand {\HII}    {\ion{H}{2}}   
\newcommand {\HeI}    {\ion{He}{1}}   
\newcommand {\HeII}  {\ion{He}{2}}   
\newcommand {\HeIII}  {\ion{He}{3}}   

\newcommand {\OI}      {\ion{O}{1}}   
\newcommand {\OIII}    {\ion{O}{3}}   
\newcommand {\OVI}    {\ion{O}{6}}   

\newcommand {\CIII}   {\ion{C}{3}}  
\newcommand {\CIV}    {\ion{C}{4}}

\newcommand {\NV}     {\ion{N}{5}}

\newcommand {\SiIV}   {\ion{Si}{4}}

\newcommand {\SiII}   {\ion{Si}{2}}

\newcommand {\kms}    {km~s$^{-1}$}

\newcommand {\FUSE}  {{\it FUSE}} 
\newcommand {\HST}  {{\it HST}}

\newcommand {\etal}  {et~al.}

\slugcomment{Accepted for publication in The Astrophysical Journal}
\shorttitle{\HeII\ Reionization Epoch}
\shortauthors{Shull, France, Danforth \etal}

\begin{document}

\title{Hubble/COS Observations of the Quasar HE~2347$-$4342:  Probing the Epoch of  \HeII\ Patchy
Reionization at Redshifts $z = 2.4-2.9$ }
\author{J. Michael Shull, Kevin France, Charles W. Danforth, Britton Smith}
\affil{CASA, Department of Astrophysical and Planetary Sciences, University of Colorado, 389-UCB, 
Boulder, CO 80309; michael.shull@colorado.edu, kevin.france@colorado.edu, charles.danforth@colorado.edu, britton.smith@colorado.edu} 
\author{and Jason Tumlinson}
\affil{Space Telescope Science Institute, Baltimore, MD 21218; tumlinson@stsci.edu}

\begin{abstract}

We report ultraviolet spectra of the high-redshift ($z_{\rm em} \approx 2.9$)  quasar, HE~2347$-$4342,
taken by the Cosmic Origins Spectrograph (COS) on the {\it Hubble Space Telescope} (\HST).  
Spectra in the G130M (medium-resolution, 1135--1440~\AA) 
and G140L (low-resolution, 1030--2000~\AA) gratings exhibit patchy Gunn-Peterson
absorption in the 303.78~\AA\  \Lya\ line of He~II  between $z =$ 2.39--2.87 (G140L) and 
$z =$ 2.74--2.90 (G130M).   With COS, we obtain better spectral resolution, higher-S/N, and 
better determined backgrounds than previous studies, with sensitivity to abundance fractions 
$x_{\rm HeII} \approx 0.01$ in filaments of the cosmic web.  The \HeII\ optical depths from COS 
are higher than those  with the {\it Far Ultraviolet Spectroscopic Explorer} (\FUSE)
and range from  $\tau_{\rm HeII} \leq 0.02$ to $\tau_{\rm HeII} \ge 5$, with a slow recovery in 
mean optical depth to $\langle \tau_{\rm HeII} \rangle \leq 2$ at $z < 2.7$.  The \HeII/\HI\ optical-depth 
ratio varies ($\eta \approx$ 10--100 for $2.4 < z < 2.73$, and $\eta =$ 5--500 for $2.75 < z < 2.89$) 
on scales $\Delta z \la 0.01$ (10.8~Mpc in comoving radial distance at $z = 2.8$), with numerous
flux-transmission windows between 1135--1186~\AA.  
The \HeII\ absorption extends to 1186.26~\AA\  ($z = 2.905$), including associated absorbers 
with $z_{\rm abs} \approx z_{\rm QSO}$ and minimal ``proximity effect"  of flux transmission at  
the  \HeII\ edge.    We propose a QSO systemic redshift $z_{\rm QSO} = 2.904\pm0.002$, some 
$\Delta z = 0.019$ higher than that derived from \OI\ $\lambda1302$ emission. Three long 
troughs (4--10~\AA\ or 25--60 Mpc comoving distance) of strong \HeII\ absorption between 
$z = 2.75-2.90$ are uncharacteristic of the intergalactic medium if \HeII\ reionized at 
$z_r \approx 3$.  Contrary to recent indirect estimates ($z_r = 3.2 \pm 0.2$) from \HI\ optical 
depths, the epoch of  \HeII\ reionization may extend to $z \la 2.7$.    \\
\end{abstract}



\section{Introduction}

The epoch of reionization  in hydrogen has become a topic of considerable interest
(Barkana \& Loeb 2001; Fan, Carilli, \& Keating 2006; Meiksin 2009; Furlanetto \etal\ 2009) 
as a probe of the transition from neutral to ionized hydrogen in the intergalactic medium (IGM).
This transition occurred somewhere between redshifts $z =$ 6--12, marking the exit from the 
cosmic ``dark ages",  
beginning at the time when the first stars and galaxies formed at redshifts $z > 30$ (Tegmark \etal\ 
1997;  Ricotti, Gnedin, \& Shull 2002; Trenti \& Shull 2010).   Helium underwent similar reionization  
from \HeII\ to \HeIII\ (that is, from He$^{+}$ to He$^{+2}$) at $z = 2.8 \pm 0.2$ 
(Reimers \etal\ 1997; Shull \etal\ 2004), most likely mediated by the harder ($E \geq 54.4$ eV) 
radiation from quasars and other active galactic nuclei (AGN).  With a 4 ryd ionization potential,  
He$^+$ is harder to ionize than H$^0$,  and He$^{+2}$ recombines 5--6 times faster than H$^+$ 
(Osterbrock \& Ferland 2006; Fardal, Giroux, \& Shull 1998).   For these reasons, 
and the fact that most hot stars lack strong 4 ryd continua, 
it is believed that AGN are the primary agents of  \HeII\ reionization.   Ionization models 
find that, owing to its resilience,  He$^+$ is much more abundant than H$^0$,
with predicted column-density ratios $\eta \equiv$ N(\HeII)/N(\HI) $\approx$ 50--100 
(Miralda-Escud\'e \etal\ 1996; Fardal, Giroux, \& Shull 1998).

Determining when and how the universe was ionized has been an important question 
in cosmology for decades (Gunn \& Peterson 1965; Sunyaev 1977).   Although recent 
progress has led to quantitative constraints,  we still do not know whether galaxies 
are the sole agents of hydrogen reionization, and the epoch of reionization remains uncertain. 
We {\it do} know that hydrogen reionization in the IGM is complete by $z \approx 6$, based on the 
rapid evolution between redshifts $z = 6.2$ and $z \approx 5$ of \Lya\ Gunn-Peterson (GP) 
absorption from neutral  hydrogen along lines of sight to QSOs (Fan \etal\ 2002; Gnedin 2004).  
The optical depth of the cosmic microwave background (Komatsu \etal\ 2010) sets a $3 \sigma$ 
constraint on instantaneous reionization at $z_r > 6.5$ and a $1 \sigma$ confidence interval of 
$z_r = 10.5 \pm 1.2$.  A limit  $z_r  \geq 7$ is set  by the detection of \Lya\ emitters at $z = 6.5$ 
(major evolution in the observed luminosity function would be expected if neutral gas were present 
around the emitters).  
These observations are in line with $\Lambda$CDM simulations (Trac \& Cen 2007;  Cen 2003;
Gnedin 2004) and ionization-front (I-front) models (Venkatesan, Tumlinson \& Shull 2003; 
Bolton \& Haehnelt 2007; Shull \& Venkatesan 2008).  The current data leave open several 
hydrogen ionization scenarios, some involving simple reionization at $z_r \approx 10$ and others 
with more complex ionization histories that depend on the star-formation rate at $z = 7-20$ and 
the transition from Pop~III to Pop~II stars.  Numerical models of the reionization process often follow 
the expansion and overlap of I-fronts, but the details are complicated by geometric uncertainties in 
the ionizing sources, higher recombination rates in denser gas, and radiative transfer of Lyman 
continuum (LyC) photons.   In particular, the gas is highly structured in both density and temperature.

The \HeII\ reionization epoch ($z_r$) and ionization processes are also poorly characterized.   The thermal 
evolution of the IGM includes many subtle effects, with degeneracies arising from fluctuations in density, 
temperature, and ionization fraction, as well as spectral hardness of the ionizing sources.  Previous
estimates of $z_r \approx 3$ were based on indirect measurements of temperature, ionization, and average
optical depth of \Lya\ lines of \HI,  and on apparent shifts in \SiIV/\CIV\ absorption-line ratios arising from 
changes in the 4-ryd  continuum radiation field as \HeII\  was ionized to \HeIII.  Early estimates of the IGM 
temperature evolution and equation of state from \HI\ line widths 
suggested a change at $z \approx 3$ (Ricotti, Gnedin, \& Shull 2000;  
Schaye \etal\ 2000).  Theuns \etal\  (2002) analyzed the average optical depth of (\HI) \Lya\  lines in 1061 
quasar spectra, using low-resolution data from the Sloan Digital Sky Survey (Bernardi \etal\ 2003).  From 
a weak dip in $\tau_{\rm eff}(z)$ at $z \approx 3.1$, these authors suggested that \HeII\ reionization started 
at redshift $z \approx 3.4$ and lasted until $z \approx 3.0$.  Several recent studies with higher-resolution 
spectroscopic data have not confirmed the $\tau_{\rm eff}$ feature (McDonald \etal\ 2005;  Kim \etal\ 2007).
Faucher-Giguere \etal\ (2008) found a feature in $\tau_{\rm eff}$(\HI) near $z = 3.2$, in a sample of
86 high-resolution, high-S/N quasar spectra taken with Keck (ESI, HIRES) and Magellan (MIKE).  However,
they question the reionization explanation of the depth and narrow redshift extent.  They also note that
at least three physical effects could produce a feature in in $\tau_{\rm eff}(z)$:  changes in IGM
temperature, electron number density, and ionizing background.   Recent numerical simulations 
(Bolton \etal\  2009; McQuinn \etal\ 2009; Lidz \etal\ 2010) also question whether \HeII\  reionization is 
responsible for the proposed feature in $\tau_{\rm eff}(z)$ at $z \approx 3.2 \pm 0.2$.   

The most direct estimates of \HeII\ reionization come from far-ultraviolet spectra of \HeII\ 
Lyman-$\alpha$ absorption toward high-redshift AGN.  
Because of strong Galactic interstellar absorption in the \HI\ Lyman continuum, the \HeII\ $\lambda 304$ 
line is only observable in the far ultraviolet, at redshifts $z \geq 2$ when the line shifts longward 
of 912~\AA.  A small number of bright AGN at $z =$ 2.9--3.3 have been observed with the {\it Hubble 
Space Telescope} (\HST) and the {\it Far Ultraviolet Spectroscopic Explorer} (\FUSE).   These data 
show moderate optical depths in the \HeII\  line, ranging up to $\tau_{\rm HeII} = 3-4$, with a gradual 
recovery of transmission ($\tau_{\rm HeII} < 1$) at $z < 2.7$.  
The best \HeII\ absorption data have been acquired toward three high-redshift AGN:  
Q0302$-$003 (Jakobsen \etal\ 1994; Hogan, Anderson, \& Rugers 1997; Heap \etal\ 2000);  and
HE~2347$-$4342 (Reimers \etal\ 1997; Kriss \etal\ 2001; Smette \etal\ 2002; Shull \etal\ 2004;  
Zheng \etal\ 2004); and HS~1700$+$6416 (Davidsen, Kriss, \& Zheng 1996; Fechner \etal\ 2006).    
From an analysis of the \HeII\ optical-depth evolution in moderate-resolution FUSE data toward 
HE~2347$-$4342,  Shull \etal\ (2004) suggested that the reionization epoch of He$^+$ occurred 
at $z_r = 2.8 \pm 0.2$.  

Observations of  \HeII\ with {\it HST} and {\it  FUSE} also constrain the thermodynamic state of 
the IGM and the spectra of ionizing sources. For example, the high abundance ratio,
N(He~II)/N(H~I) $\approx$ 50--100, observed toward HE~2347$-$4342 and HS~1700$+$6416 
(Kriss \etal\ 2001; Smette \etal\ 2002; Shull \etal\ 2004; Zheng \etal\ 2004; Fechner \etal\ 2006) 
requires a soft ionizing radiation field, $F_{\nu} \propto \nu^{-\alpha_s}$, with mean spectral index 
$\langle \alpha_s \rangle \approx 1.8$ (Fardal \etal\ 1998).  As observed in the QSO rest-frame, 
the EUV (1--2 ryd) spectral indices of quasars exhibit a wide range (Telfer \etal\ 2002; Scott \etal\ 2004).  
More intriguing are the spectroscopic observations of \HeII\ and \HI\ taken by  {\it FUSE}, the Keck Telescope, 
and the Very Large Telescope (VLT).  Kriss \etal\  (2001) and Shull \etal\ (2004) report variations in 
$\eta = 20-200$ over 2--10 Mpc scales ($\Delta z \approx  0.002-0.01$) which imply spatial fluctuations 
in the radiation field at the ionization edges of  \HI\ (1~ryd) and \HeII\ (4~ryd).  Most of these variations 
probably arise in the 4~ryd continuum, whose photons have greater penetrating power.  These flux 
differences at 1 ryd and 4 ryd are then amplified by absorption and reprocessing by the IGM 
(Haardt \& Madau 1996; Fardal \etal\ 1998).   

In this paper, we present new, high-quality, far-UV spectroscopic observations of intergalactic 
\HeII\ absorption toward  HE~2347$-$4342, using the Cosmic Origins Spectrograph (COS) on 
\HST\ (Green \etal\ 2010;  Osterman \etal\ 2010).  
This target is of considerable interest as a bright ($V = 16.1$, $z \approx 2.9$) 
continuum source, previously observed by the \FUSE, Keck, and VLT spectrographs to study 
intergalactic \Lya\ absorption in both \HI\ and \HeII.   HE~2347$-$4342 is the first of three AGN 
targets scheduled for COS guaranteed-time observations of the \HeII\ reionization epoch.  
The data were acquired with both the moderate-resolution G130M grating ($R \approx 18,000$, 
$\lambda = 1135-1440$~\AA) and the low-resolution G140L grating ($R \approx 1500$, 
$\lambda = 1030-2000$~\AA).  In the 303.78~\AA\  line of \HeII, these wavelength bands allow 
us to probe redshifts  down to $z_{\rm HeII}  =  2.735$ (G130M) and $z_{\rm HeII} \approx  2.39$ 
(G140L).  The COS data enable many improvements compared to previous studies.  
First, the high far-UV throughput of the COS/G130M grating provides higher signal-to-noise (S/N) 
and better photometric accuracy.  Second, the low background of the COS detectors allows us to 
detect very weak flux-transmission through the IGM, characterize low flux levels in \HeII\ absorption 
troughs, and probe regions of high optical depth, $\tau_{\rm HeII} \ga 5$.   Finally,  low-resolution 
G140L spectra (1000-2000 \AA) allow us to study the recovery of the  \HeII\ optical depth at 
$z < 2.7$ and characterize the continuum longward of the \HeII\ edge.

In \S2 we discuss the observations and data reduction techniques for both G130M  and G140L gratings.   
In \S3 we display the fitted quasar continuum, $F_{\lambda} \propto \lambda^{-3.0 \pm 0.1}$, which we 
use to extrapolate below the \HeII\ edge at $1186.2$~\AA.   From this continuum and the transmitted
fluxes, we derive \HeII\ optical depths in the \Lya\ forest and analyze the fluctuating absorption.  
We observe minimal effects of local photoionization expected from proximity to this very luminous QSO.   
In the \HeII\  \Lya\ forest at $2.4 < z < 2.9$, we see narrow (2--5~\AA, 12--30 Mpc of comoving radial distance) 
windows of partial flux transmission, which we interpret as ``patchy \HeII\ reionization", interspersed with 
broad \HeII\  absorption troughs (4--10~\AA\ intervals, 25--60 Mpc) with high optical depths, 
$\tau_{\rm HeII} \geq 5$.   At $\lambda < 1130$~\AA\  ($z_{\rm HeII} < 2.72$) the mean opacity drops to 
$\langle \tau_{\rm HeII} \rangle \la 2$, consistent with the overlap of  the \HeII\  I-fronts.  Because \HeII\ is 
much more abundant than \HI, its opacity  through the cosmic web remains high.

In \S4 we summarize our observations and their implications for the \HeII\ reionization epoch.  
With reasonable assumptions about the physical state of the IGM at $z = 2-3$, borne out by studies
of  \HI\ and \HeII\ \Lya-forest absorption and cosmological simulations,  we can probe the \HeII\ 
reionization epoch down to fractional ionization $x_{\rm HeII} \geq (0.01)(0.1/\delta_H)(\tau_{\rm HeII} /5)$ 
in underdense regions ($\delta_H = 0.1-1$) relative to the mean baryon density of the cosmic web.  
This ionization level provides much greater sensitivity to conditions at $z \approx 2.8$ than possible 
with \HI\ at $z \approx 6$ (Fan \etal\ 2006).   
Coupled with galaxy surveys, cosmological simulations, and photoionization modeling, these UV
spectra will help us understand the propagation and overlap of the helium ionization fronts.


\section{$HST$-COS Observations and Data Reduction}

\subsection{Wavelength Scale and Fluxes} 

HE\,2347$-$4342 was observed with the short-wavelength, medium-resolution far-UV 
mode of HST-COS (G130M) on 2009 November 05 for a total of 28,459~s.   Descriptions
of the COS instrument and on-orbit performance characteristics are in preparation 
(Green \etal\ 2010;  Osterman \etal\ 2010).   In order to achieve continuous spectral coverage 
across the  G130M bandpass (1135~\AA\  $\la \lambda \la$ 1440~\AA) and to minimize fixed-pattern 
noise, we made observations at two central wavelength settings (1291~\AA\ and 1300~\AA) 
with four focal-plane offset locations in each grating setting (FP-POS = 1, 2, 3, 4).
This combination of grating settings ensures the highest signal-to-noise observations at the 
shortest wavelengths available to the G130M mode ($z_{\rm HeII} \approx 2.73-3.00$) at a
resolving power of $R  \equiv (\lambda/\Delta \lambda) \approx 18,000$ 
($\Delta v_{\rm FWHM} \approx 17$~\kms).  

In addition, COS observed HE\,2347$-$4342 in the low-resolution G140L mode for
11,558~s on 2009 November 06.  The G140L Segment-A data provide some overlap 
with the G130M observations between 1240--1440~\AA,
and they extend our low-resolution spectral coverage ($R \approx 1500$) out to 
$\lambda \approx 2000$~\AA.  From G140L segment-B observations and custom 
data-processing steps described in Section 2.2, we extend our reliable spectral coverage 
over part of the G140L bandpass, between 1030--1175~\AA.  Table~1 lists the relevant 
COS observational parameters for the data sets used in this work.


\begin{deluxetable}{cccccc}
\tabletypesize{\footnotesize}
\tablecaption{HE~2347$-$4342 Observing Log }
\tablewidth{0pt}
\tablehead{
\colhead{Date}  & \colhead{COS Mode} &   \colhead{$\lambda_{\rm c}$ (\AA)}   & \colhead{FP-POS} &
     \colhead{Exp} & \colhead{$T_{\rm exp}$ (s)}     
     }
\startdata	
2009 Nov 05	  &  	G130M 	& 	1291~\AA\    &   1,2,3,4   &    9    &   16535	      \\
2009 Nov 05	  &  	G130M 	& 	1300~\AA\    &   1,2,3,4   &    9    &   11924       \\
2009 Nov 06	  &  	G140L 	& 	1230~\AA\    &   1,2,3,4   &    4    &   11558       
 \enddata


\end{deluxetable}


All observations were centered on the QSO target, HE\,2347$-$4342 
at J2000 coordinates (R.A. $ = 23^{\mathrm h} 50^{\mathrm m} 34.23^{\mathrm s}$,
Decl. = $-43\arcdeg 25\arcmin 59.80\arcsec$).  COS performed NUV imaging target
acquisitions with the MIRRORA/PSA mode, yielding good centering in the
COS aperture to maximize throughput and resolving power.
The G130M data were processed with the COS calibration pipeline,
{\sc CALCOS}\footnote{See the \HST\ Cycle 18 COS Instrument Handbook for more
details:\\ {\tt http://www.stsci.edu/hst/cos/documents/handbooks/current/cos\_cover.html}}
v2.11b, and combined with a custom IDL coaddition procedure. 
The analog nature of the COS micro-channel plate makes it susceptible to temperature changes. 
Electronically injected pulses (stims) at opposite corners of the detector allow for the tracking 
and correction of any drift (wavelength zero-point) and/or stretch (dispersion solution) of the 
recorded data location as a function of temperature.  At the beginning of the observations of 
HE~2347$-$4342, the detector proved to be substantially colder than the nominal operating 
temperature of the micro-channel plate.  This allowed one of the electronic stim-pulses  to
accumulate at the bottom of the detector, compromising the wavelength solution for the first 
four  G130M exposures.   The detector warmed to the normal operating range for the remaining 
G130M observations.  We reprocessed the earlier exposures using extrapolated stim positions 
to ensure a consistent wavelength solution over the course of the observations.  
The G140L data were unaffected.  

Flat-fielding, alignment, and coaddition of the processed COS/G130M exposures were carried out 
using IDL routines developed by the COS GTO team specifically for COS FUV 
data\footnote{See {\tt http://casa.colorado.edu/$\sim$danforth/costools.html} for our
coaddition and flat-fielding algorithm and additional discussion.}.
First, the data were corrected for fixed-pattern instrumental features.  Although attempts at a true 
``flat-fielding'' of COS data show promise, the technique is not yet sufficiently robust to improve 
data of moderate S/N.  
The quantum efficiency is improved by the presence of a series of wires, called the quantum-efficiency 
grid, placed above the detector.   These wires create shadows in the spectrum that are removed during 
data reduction.  An ion-repellor grid reduces the background rate by preventing low-energy thermal ions 
from entering the open-faced detector.  We are able to correct the narrow shadows of 
$\sim$15\% opacity arising from the grid wires.   A one-dimensional map of grid-wire opacity for each
detector was shifted from detector coordinates into wavelength space and divided from the flux and error 
vectors.  Exposure time in the location of grid wires was decreased by $\sim0.7$, giving these pixels
less weight in the final coaddition.  We also modified the error vector and local exposure time at the edges
 of the detector segments to underweight the flux contributions from these regions.  With a total of eight 
 different wavelength settings in the final co-addition, any residual instrumental artifacts from 
 grid-wire shadows and detector segment boundaries should have little effect on the final spectrum.

We performed two sets of wavelength-scale shifts.  First, the G130M exposures were aligned with 
each other and interpolated onto a common wavelength scale.  Then, we transferred the G130M 
wavelengths to a ground-based reference frame of the \HI\ data, taken by D. Reimers with the 
VLT/UVES (Ultraviolet Visual Echelle Spectrograph) in ESO program 071.A-0066.  These data 
were used in the \HeII\ analysis paper of \FUSE\ data (Shull \etal\ (2004).     
One exposure in each COS detector was picked as a wavelength reference, and the remaining 
exposures were cross-correlated with it.  The wavelength region of cross-correlation for each case 
was picked to include a strong interstellar absorption feature, although shifts were typically 
a resolution element or less.  Residual errors on the target centering within the COS aperture 
caused a systematic shift between the strong interstellar lines and their expected LSR velocities.  
The COS wavelength solution is relatively accurate, but it may suffer from linear offsets from 
an absolute frame.  Since our later analysis involves comparing the rapidly-varying \HI\ (optical) 
and \HeII\  (UV) profiles, wavelength alignment is critical.  The only absorption features in the 
normalized VLT data clearly not associated with the \Lya\  forest are a pair of narrow absorbers 
at 4787.12 \AA\ and 4795.08 \AA,  consistent in relative strength and separation with the \CIV\ doublet 
at 1548.195 and 1550.770~\AA,  redshifted to $z = 2.0921$.  At this redshift, COS/G130M covers 
384--466~\AA.  The  \CIII\  $\lambda 386.20$ line at 1194.1~\AA\  in the COS data corresponds to 
the \CIV\ doublet in the optical data, with similar line profiles.   
We therefore shift the COS data by $+0.07$~\AA\ (approximately one G130M 
resolution element) to align with the optical spectra.

The flux at each position was taken to be the exposure-time weighted mean of flux.
Since exposure time was reduced in certain wavelength locations, as noted above, pixels near 
detector edges and where grid-wire shadows were removed received less weight than those in 
less suspect locations.  The combined G130M data show $S/N\sim 25$ per 7-pixel resolution 
element, sufficient to detect narrow absorption features near 1200~\AA\ down to equivalent widths
$W_{\lambda} \ga 10$ m\AA\ at $4\sigma$ significance.
The G140L observations were originally processed with {\sc CALCOS} 2.11b.  However, the 
$\lambda<1200$~\AA\ data from this processing is not appropriate for scientific analysis. 
We describe our custom processing in Section 2.2.


\begin{figure*}
  \epsscale{1.1}
  \plotone{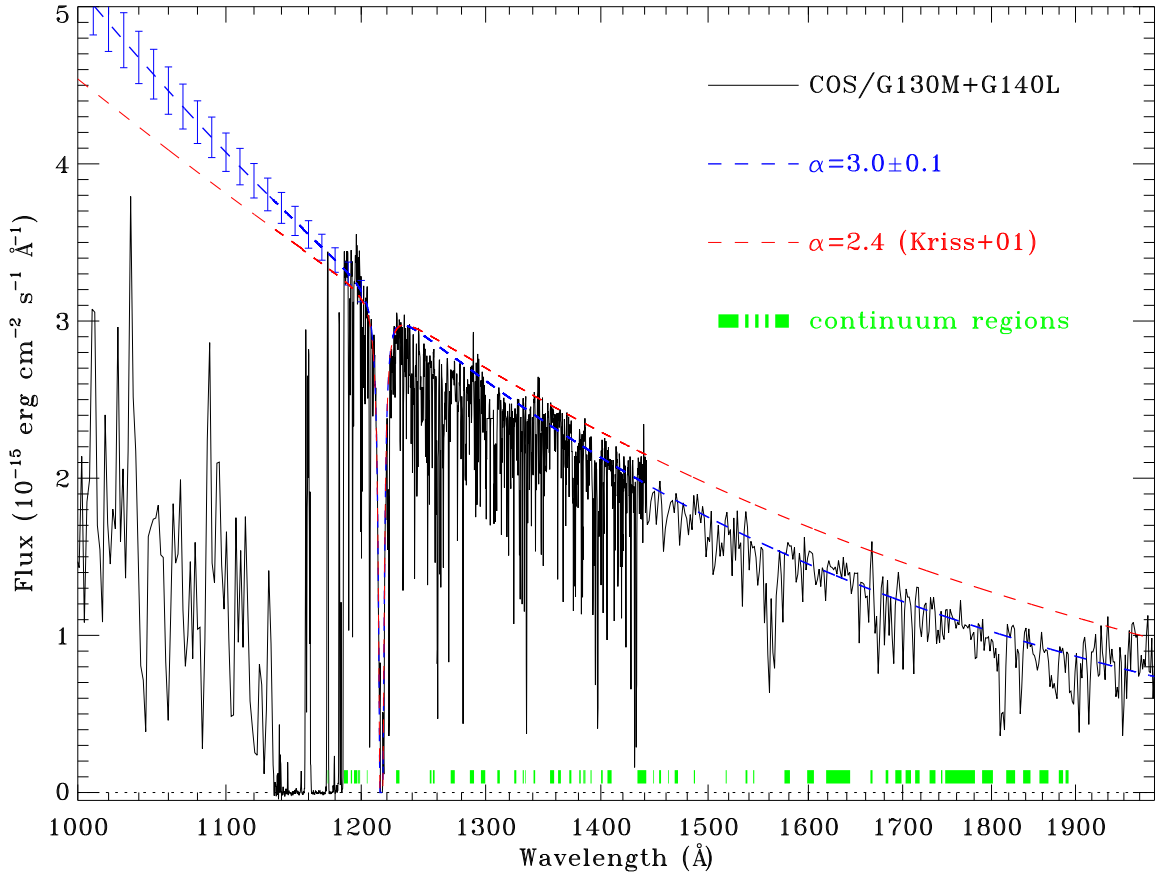}    
   \caption{Overview of the COS/G130M and G140L data.  Data are smoothed by 3 resolution elements
       and shown as observed  (uncorrected) flux. The de-reddened data were fitted to a power law, 
       $F_{\lambda} = F_0 (\lambda / \lambda_0)^{-\alpha_s}$, with $\alpha_s = 3.0 \pm 0.1$.
       Error bars below $\lambda_0 = 1215$~\AA\  reflect  $\pm2$\% uncertainty in continuum level and
      $\pm 0.1$ uncertainty in $\alpha_s$.   This continuum is extrapolated below the He\,II edge (1186~\AA)
      to derive He\,II optical depths.  Red curve shows  continuum fit of Kriss \etal\ (2001) with $\alpha_s = 2.4$.
     Our steeper fit arises primarily from the higher quality COS data (1200--2000 \AA) and, to lesser 
     extent, from our larger adopted extinction, $E(B-V) = 0.022$ rather than 0.014, extrapolated to the far-UV. }
\end{figure*}



\begin{figure*}
   \epsscale{1.2} 
   \plotone{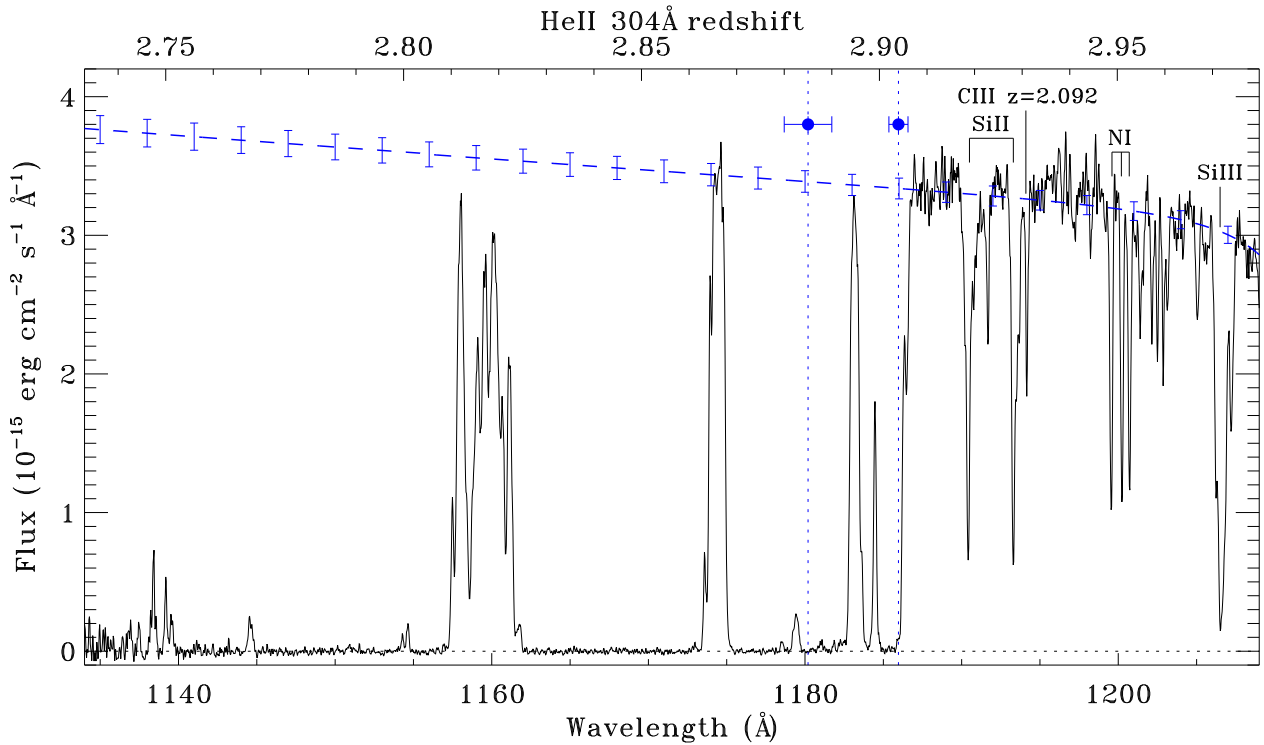}    
   \caption{Detail of the G130M data showing He\,II absorption and transmission windows. 
   The two proposed QSO systemic redshifts ($z_{\rm QSO}  = 2.885$ and 2.904) and our extrapolated 
   continuum are marked.  Milky Way interstellar lines of N\,I, Si\,II, Si\,III appear longward of 1190~\AA, 
   and strong He\,II absorption is seen shortward of 1186.26 \AA.  We use the C\,III ($z = 2.0921$) IGM 
   absorption line to align the COS and VLT spectra.  Windows of flux transmission appear at 1138, 1144, 
   1154, 1160, 1174,  1179, 1183, and 1184~\AA.   At 1174~\AA\, we see essentially 100\% transmission, 
   and yet we also observe three troughs of strong He\,II absorption  ($\tau_{\rm HeII} \geq 5$) over 
   large redshift intervals (and comoving radial distances) between $z = 2.751-2.807$ (61 Mpc), 
   $z = 2.823-2.860$ (39 Mpc), and $z = 2.868-2.892$ (25 Mpc). }
\end{figure*}
  


\begin{figure*}
  \epsscale{1.0}
  \plotone{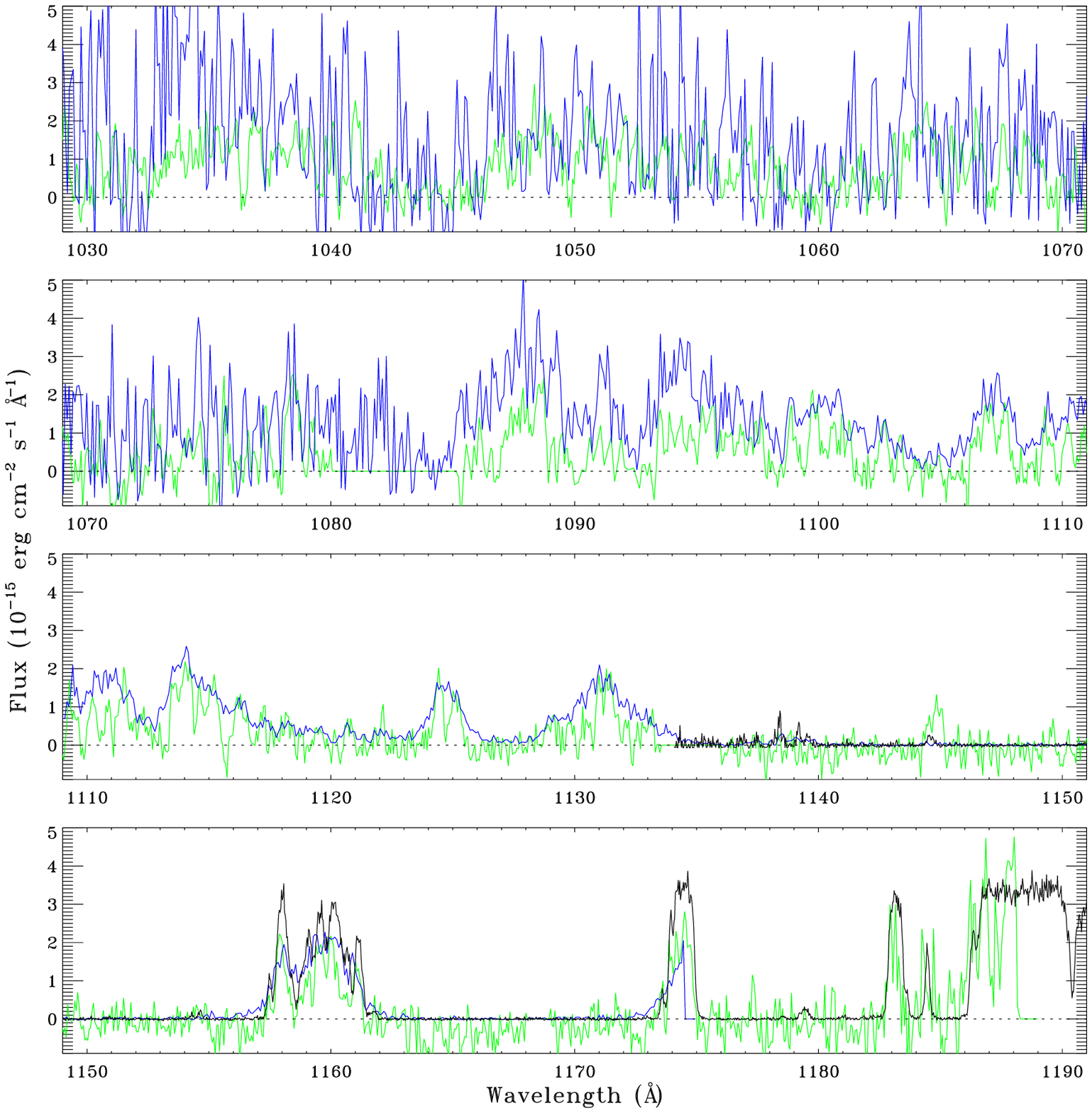}    
   \caption{Comparison of the COS/G130M (black), COS/G140L (blue), and coadded \FUSE/LiF1+LiF2 data 
   (green).  The COS/G130M data are binned by 3 pixels (half a resolution element).  The \FUSE\ data have 
   been binned to 
   one resolution element, and COS/G140L data are presented as observed (unbinned).  The \FUSE\ 
   background solutions appear to be adequate at 1110-1150 \AA, but poor between 1160--1187 \AA.  
   The flux and wavelength solutions for the two instruments are reasonably consistent, providing 
   confidence in using G140L data at $\lambda >1030$~\AA.  }
\end{figure*}


Figure 1 shows the G130M/G140L continuum from 1200~\AA\ to 2000~\AA, which we have fitted 
and extrapolated below the \HeII\ edge at 1186.2~\AA.   The G140L wavelengths were shifted to 
match G130M, with no corrections needed to match G130M and G140L fluxes.  These curves were 
de-reddened using color excess, $E(B-V) = 0.022$, corresponding to column density 
$\log N_{\rm HI} = 20.06$ derived by fitting interstellar \Lya\ absorption.  We adopted a standard 
interstellar ratio, N$_{\rm HI} = (5.3  \times 10^{21}$~cm$^{-2}$~mag$^{-1}) E(B-V)$, from 
Shull \& Van Steenberg (1985) and Diplas \& Savage (1994) and used the UV selective extinction 
curve of Fitzpatrick \& Massa (1999).  Our derived extinction, $E(B-V) = 0.022$, lies intermediate 
between previous values of $E(B-V) = 0.0387$ (Reimers  \etal\ 1997) and $E(B-V) = 0.014$ 
(Smette  \etal\ 2002).  Figure 2 shows the G130M spectrum ($\lambda < 1208$~\AA) with prominent 
\HeII\ absorption troughs, numerous flux-transmission windows shortward of the \HeII\ edge, and 
the two proposed QSO systemic redshifts, $z_{\rm QSO} = 2.885 \pm 0.005$ (Reimers \etal\ 1997) 
and $z_{\rm QSO} = 2.904 \pm 0.002$ (this paper).  This issue is discussed further in Section 3.1. 

The strong  \HeII\ features are listed in Table 2, with the \HeII\ optical depths calculated from
the depth of flux transmission, $\tau_{\rm HeII} =  \ln (F_0 / F_{\lambda})$, relative to the
extrapolated continuum $F_0$.    Uncertainties in $\tau_{\rm HeII}$ were computed as 
root-mean-square averages over redshift bins, $\Delta z$, ranging from 0.004 to 0.02 at
wavelength centroids $\lambda_c$.   We conservatively rounded errors to the nearest 0.25 in 
$\tau_{\rm HeII}$, using uncertainties arising from converting errors in flux transmission to errors 
in optical depth.  In regions of lower S/N in the G140L grating ($z \la 2.6$), we smoothed over 
two resolution elements  ($\sim 2$~\AA) to reduce the noise before the optical depth and error 
were evaluated.  The G140L-B spectra shown in Figure 1 were smoothed over three 
resolution elements.





\begin{deluxetable}{cccccc}
\tabletypesize{\footnotesize}
\tablecaption{HE~2347-4342 Sight Line He\,II Absorption\tablenotemark{a}}
\tablewidth{0pt}
\tablehead{
\colhead{$z_c$}  & \colhead{$\Delta z$}  &  \colhead{$\lambda_{c}$ (\AA)}  & 
      \colhead{$\tau_{\rm HeII}$} &  \colhead{COS Mode} & \colhead{Notes\tablenotemark{b}}  
}
\startdata	
 2.90151   &  0.00400     &  1185.20       &  $\geq$~5               &  G130M  &        $\cdots$    \\
 2.89765   &  0.00800     &  1184.03       &  $\geq$~5               &  G130M  &        $\cdots$    \\
 2.88721   &  0.00823     &  1180.86       &  $\geq$~5               &  G130M  &        $\cdots$    \\
 2.87445   &  0.01317     &  1176.98       &  $\geq$~5               &  G130M  &        Z04-3A    \\      
 2.84700   &  0.03456     &  1168.64       &  $\geq$~5               &  G130M  &        Z04-3CDE   \\
 2.77471   &  0.05267     &  1146.68       &  $\geq$~5               &  G130M  &        Z04-3HIJ   \\
 2.71324   &  0.00823     &  1128.01       &  $3.3 \pm 0.25$     &   G140L   &     $\cdots$     \\
 2.69082   &  0.01646     &  1121.20       &  $3.0 \pm 0.50$     &   G140L   &   $\cdots$     \\
 2.66414   &  0.00494   &  1113.09         &  $1.8 \pm 0.25$     &   G140L   &     $\cdots$   \\
 2.65061   & 0.00494    &  1108.98         &  $1.5 \pm 0.25$     &   G140L   &     $\cdots$   \\
 2.63785   &  0.01317     &  1105.11       &  $2.3 \pm 0.25$     &   G140L   &     $\cdots$   \\
 2.61814   &  0.00988   &  1099.12         &  $1.3 \pm 0.50$     &   G140L   &     $\cdots$   \\
 2.59804   &  0.00988   &  1093.01         &  $2.0 \pm 1.0$       &   G140L   &     $\cdots$   \\  
 2.57059   &  0.01317     &  1084.67       &  $3.5 \pm 1.0$       &   G140L   &     $\cdots$   \\
 2.55000   &  0.02008     &  1078.42       &  $1.0 \pm 0.25$     &   G140L   &     $\cdots$   \\
 2.52999   &  0.02008     &  1072.34       &  $2.5 \pm 0.50$     &   G140L   &     $\cdots$   \\
 2.49327   &  0.01975     &  1061.19       &  $2.8 \pm 1.0$       &   G140L   &     $\cdots$   \\
 2.47001   &  0.02008     &  1054.12       &  $1.5 \pm 0.50$     &   G140L   &     $\cdots$   \\
 2.43683   &  0.01811     &  1044.04       &  $1.7 \pm 0.50$     &   G140L   &    $\cdots$     
 \enddata

\tablenotetext{a}{Average He\,II optical depth $\tau_{\rm HeII}$ was computed from flux
transmission averaged over redshift bins $z_c \pm \Delta z$ at centroid wavelengths $\lambda_c$.
Errors in $\tau_{\rm HeII}$ are conservatively rounded to nearest 0.25.}
\tablenotetext{b}{These three He\,II complexes noted Z04-3$n$ were identified in 
Table 3 of Zheng et al. (2004). } 
\end{deluxetable}

\subsection{Data Processing: COS G140L Segment B}

During the Servicing Mission-4 Observatory Verification (SMOV-4) period, it was
discovered that the MgF$_{2}$/Al mirrors of $HST$ have retained approximately 80\% 
of their pre-flight reflectivity at $\lambda < 1175$~\AA\ (McCandliss \etal\ 2010).  
This result has opened the door for use of the short-wavelength response of COS, in 
particular the G140L mode, to perform spectroscopic observations at wavelengths 
inaccessible to previous \HST\ instruments ($\lambda < 1130$~\AA).
See Appendix~A of  France \etal\ (2010) for preliminary work with this mode. 
The {\sc CALCOS} pipeline processing of G140L segment B
(400~\AA\ $\la \lambda \la$ 1175~\AA) is not sufficiently
mature to produce one-dimensional spectra appropriate for scientific analysis.

We therefore processed the data with a newer version of the {\sc CALCOS}
pipeline (v2.12), employing specially created reference files for flux
and wavelength calibrations in the short-wavelength segment-B setting.
The custom reference files that were created to supplement the {\sc CALCOS} 2.12 
release were a first-order dispersion solution for $\lambda < 1150$~\AA\ and a flux 
calibration curve.  The wavelength calibration was created by fitting line centers to the
pressure-broadened Lyman series lines observed in COS calibration spectra of white 
dwarfs (McCandliss \etal\ 2010).  While G140L segment A uses a second-order
polynomial wavelength solution, the relatively small amount of
physical detector area used by G140L segment B allows a
reasonable fit with a linear solution.  We find a dispersion of
$\Delta \lambda = -79.8$ m\AA\ pixel$^{-1}$, and a long-wavelength
zero-point of 1245.26~\AA.   To align with G130M data, we shifted  
G140L-B wavelengths by $+3.9$~\AA.   For example, 1245.26~\AA\ (old G140L) 
becomes 1249.16~\AA\ (new G140L).   The G140L-A wavelengths were shifted 
by  only $-0.4$~\AA\  ($\sim 0.5$ resolution element).  

The calibration curve,  translating count rate (counts~s$^{-1}$) into flux
(erg~cm$^{-2}$~s$^{-1}$~\AA$^{-1})$ was created using the G140L segment-B 
effective area curve derived by McCandliss \etal\ (2010).  
We caution that there may still be significant uncertainty in this flux calibration.   
At $\lambda < 950$~\AA, we  should be within $\sim50$\% of the correct value.    
The accuracy is likely $\sim$20--25\% from 1025--1110~\AA\ and  $\sim$10--15\% 
from 1120~\AA\ longward.  Continued testing of G140L-B observations will refine 
this calibration.  This custom reduction produces reprocessed one- and two-dimensional
spectral data products, which can be coadded following the procedure described above 
and in Danforth \etal\ (2010).  The coadded data were used to create a low-resolution 
($R \approx 10^3$, $\Delta \lambda \sim 1$~\AA) spectrum of HE~2347$-$4342 at 
$\lambda \le 1174$~\AA.  In principle, the COS/G140L wavelength coverage extends to 
very low wavelengths ($\sim400$~\AA).  
However, the combination of low instrumental effective area, $A_{\rm eff} \sim 10$~cm$^{2}$ 
at $\lambda \la 1030$~\AA\ (McCandliss \etal\ 2010) and interstellar absorption by
atomic and molecular hydrogen limit the working wavelength coverage.  
For the \HeII\ absorption toward HE~2347$-$4342, we present G140L data from 
1030~\AA\ $\la \lambda \le$ 1174~\AA, corresponding to $z_{\rm HeII}  \approx 2.39-2.87$.
We cross-calibrated the G140L segment-B spectrum with the short-wavelength data from 
G130M, finding that the flux calibration is better than 10\% in the overlap
region.  The data were then used to extend our investigation of the \HeII\ 
ionization state in the IGM down to $z_{\rm HeII}  = 2.39$ (1030~\AA).  

Figure 3 provides a comparison among three spectrographs:  COS/G130M, COS/G140L, 
and \FUSE\ (LiF1 and LiF2 channels).  There is a fair amount of agreement in fluxes
among these instruments, although the \FUSE\ backgrounds at 1160--1187~\AA\ 
are obviously incorrect.  This comparison gives us confidence in using G140L data
down to 1030~\AA.

\section{Results and Analysis} 


\subsection{Systemic Redshift of QSO} 

The systemic redshift of HE~2347$-$4342 is of interest, both to fix the expected location
of the \HeII\ edge and to infer the metagalactic radiation field from the QSO proximity
effect (Dall'Aglio, Wisotzki, \& Worseck 2008a,b).   A significant puzzle for this sight line
is the 6~\AA\ offset between the extent of strong \HeII\ absorption (out to 1186 \AA) 
and the expected \HeII\ edge (1180 \AA) based on an uncertain QSO emission-line redshift.  
Reimers \etal\ (1997) found $z_{\rm em} = 2.885 \pm 0.005$, based on the weak 
\OI\  $\lambda1302$ emission line, while Scannapieco \etal\ (2006) quoted $z_{\rm em} = 2.871$
with no error bar.  Emission-line redshifts were also measured with UVES at the VLT/UT2 
by Dall'Aglio \etal\ (2008b), who quoted $z_{\rm em} = 2.886 \pm 0.003$ based on the 
\OI\ + \SiII\  blend and assuming a rest wavelength $\lambda_{\rm rest} = 1305.77$ for the
blend.  Given the solar abundance ratio (O/Si)$_{\odot} \approx 15$, this blend is probably 
dominated by three lines of \OI, with a statistically weighted centroid at 1303.49~\AA\  
(Morton 2003).   In total, the blend consists of three \OI\ lines decaying to the ground-state 
fine-structure levels of \OI\ [$(2p^3\,3s)\, ^{3} {\rm S}_{1} \rightarrow (2p^4)\, ^{3} {\rm P}_{2,1,0}]$ 
at  1302.17, 1304.86, 1306.03 \AA\ and two lines of \SiII\ $ 
[(3s\,3p^2) \, ^{2} {\rm S_{1/2}} \rightarrow (3s^2\,3p)\, ^{2} {\rm P}_{1/2,3/2}]$ 
at 1304.37 and 1309.28 \AA.   Dall'Aglio \etal\ (2008a) measured the redshifts of the strongest 
emission lines to be $z_{\rm em} = 2.877 \pm 0.003$ (\Lya) and $2.861 \pm 0.003$ (\CIV).   
However, there are well-known shifts in emission-line redshifts between the QSO systemic redshift 
and the ``higher ionization lines"  (e.g., \Lya, \CIII\ $\lambda$977, \CIV\ $\lambda$1549).   
For instance, 1000-3000 \kms\ offsets are commonly seen in the emission-line spectra of AGN 
(Espey \etal\ 1989; Corbin 1990).  These issues provide systematic uncertainties in determining  
$z_{\rm em}$, either from the weak \OI/\SiII\  blend or from higher ionization lines.  Thus, the 
precise QSO systemic redshift remains uncertain.  

Because of these problems with emission-line redshifts, we propose to use IGM absorption 
to determine the QSO redshift.  We suggest that the QSO systemic redshift is
$z_{\rm QSO} = 2.904 \pm 0.002$ (Figure 4) based on  the extent of the \HI\ and \HeII\ absorption.
This redshift is chosen as the midpoint between the centroid of the strong absorption feature
seen in both \HI\ and \HeII\  and the red edge of \HeII\ absorption.  
Figure 4 provides an overview of the COS/G130M spectrum from 1100--1190~\AA, 
together with a blowup of the region from 1173~\AA\ to the \HeII\ edge near 1186~\AA.  
We see absorption extending well beyond the wavelengths (1180.2--1180.5~\AA)
corresponding to the QSO emission-line redshifts, $z = 2.885 \pm 0.005$ and $2.886 \pm 0.003$, 
suggested by Reimers \etal\  (1997) and Dall'Aglio \etal\ (2008b) respectively. The absorption
between 1181--1186~\AA\ suggests that the QSO actually lies at $z \approx 2.90$.  Interestingly, we 
see no ionization effects from proximity to the QSO, a topic discussed further in Section 3.6.


 \begin{figure*}
   \epsscale{1.2}
    \plotone{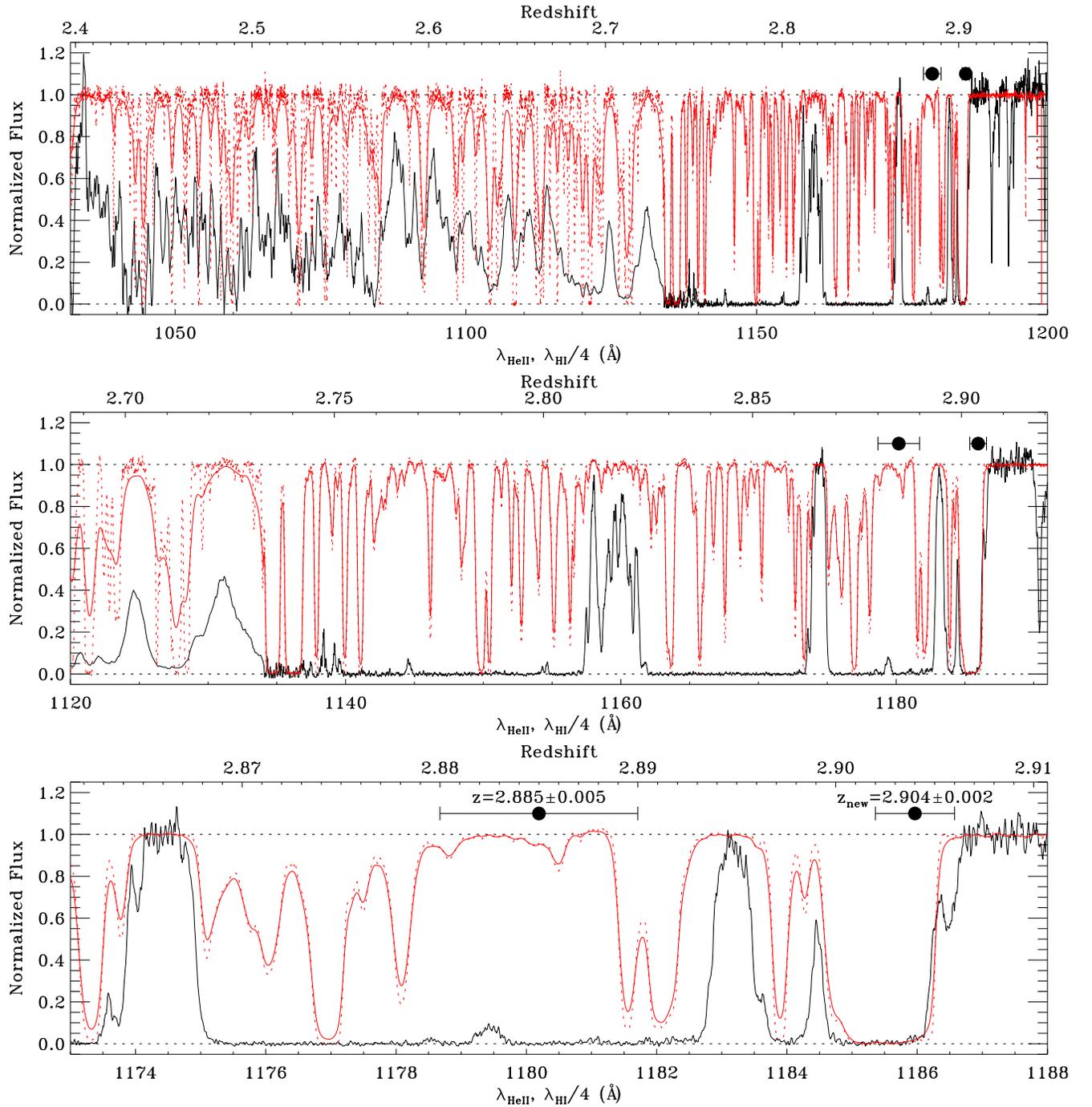}    
    \caption{(Top Panel) Normalized COS/G130M and G140L flux (black) and VLT flux (red) 
    vs.\ redshift ($z = 2.4 - 2.9$, top legend) and wavelength (bottom legend).  
    H\,I wavelengths are divided by 4 for comparison with He\,II, and two proposed QSO 
    redshifts are noted by filled black circles:  $z_{\rm em} = 2.885 \pm 0.005$ (Reimers \etal\ 1997)
    and $z = 2.904 \pm 0.002$ (this paper).   Lower two panels show VLT/UVES data ($R = 44000$) 
    from D. Reimers, convolved with COS line spread function to match COS resolution:  
    $R = 18,000$ ($\lambda > 1134$~\AA) and $R = 1800$ ($\lambda < 1134$~\AA).  Unconvolved 
    H\,I  data are shown by dotted red curves.   Troughs of strong He\,II absorption (1175.2--1179.1~\AA, 
    1179.5--1182.5~\AA) are separated by small windows of flux transmission.   Features at 
    1181-1186~\AA\ are associated absorbers (Reimers \etal\ 1997), but with insufficient column 
    density to block AGN ionizing flux.  }
\end{figure*}


The \HeII\ absorption profile shows an absorption feature at 1186.5~\AA,  with optical depth
$\tau  \approx 0.7$, equivalent width 95~m\AA, and FWHM = 0.14~\AA.  This may be proximate 
\HeII\ absorption or absorption from an intervening system.  
The spectrum near 1186.5~\AA\ shows a clear flux minimum and recovery toward shorter 
wavelengths typical of weak absorption lines.  A distinct \HeII\ absorber would appear as a broader 
``shelf" on the side of the $\tau_{\rm HeII}  > 1$ trough.  We do not make a specific line identification 
for the feature.  It does not seem to be  low-redshift \Lyb\  nor any of the 
usual IGM metal-line systems (Danforth \& Shull 2008).  

Reimers \etal\ (1997) attribute  the \HI\ and \HeII\ absorption at $z > 2.885$ to a multicomponent system 
of associated absorbers between 1181--1186~\AA\ (redshifts $z = 2.891$ to $z = 2.904$).  Many of these 
systems have strong metal lines of \CIV, \NV, and \OVI, and a few have anomalous ratios of  \HeII\ and \HI\ 
\Lya\ absorption.  These observations suggest that the gas might be exposed to high fluxes of
ionizing radiation and affected by nucleosynthetic anomalies in nuclear outflows.  Smette \etal\
(2002) and Fechner \etal\ (2004) both noted evidence for  ``line-locking" in \CIV\ and \NV\ (although
the latter group found no statistical evidence compared to simulated line lists).  Line-locking occurs
when two absorbers are separated by a velocity equal to the separation of the two doublet lines, resulting
from radiation pressure on material physically near the QSO (Weymann, Carswell, \& Smith 1981).  These 
radiative forces will drive gas outward from the QSO, whereas the redshifted associated absorbers would 
have infall velocity $\sim1500$ \kms\ relative to the QSO if $z_{\rm QSO} = 2.885$.  Thus, our proposed 
redshift, $z_{\rm QSO} = 2.904$, provides a more plausible explanation.

The uncertainty in QSO systemic redshift illustrates the need for infrared observations.  
Unfortunately, the redshifted hydrogen Balmer lines fall in difficult spectral bands (H$\alpha$ at 
2.563~$\mu$m, H$\beta$ at 1.898~$\mu$m), as does the [O~II] $\lambda 3727$ doublet 
at 1.455 and 1.456 $\mu$m.  The forbidden lines of [O~III] $\lambda 5007, 4959$ occur at more 
promising wavelengths, 1.955~$\mu$m and 1.936~$\mu$m.  Thus, infrared spectroscopic searches 
for the [O~III] forbidden lines may offer the best opportunity to resolve the issue of $z_{\rm QSO}$.


\subsection{Expected He~II Gunn-Peterson Absorption}

\noindent
The standard GP optical-depth formula (Fan \etal\ 2006) for species ($i$ = \HI\ or \HeII) is 
\begin{equation}
  \tau_i(z) = \left( \frac {\pi e^2}{m_e c} \right) \frac {\lambda_i  \, f_i \, n_i } {H(z)}  \;  ,
  \end{equation}
where $(\pi e^2/m_e c) = 2.654 \times 10^{-2}$~cm$^2$~Hz is the integrated classical oscillator
cross section, $f_i = 0.4162$ is the absorption oscillator strength for \Lya\ transitions of  \HI\ and \HeII\ 
at rest wavelengths $\lambda_i = 1215.67$~\AA\ and 303.78~\AA, and $n_i$ is their number density 
(cm$^{-3}$).   The Hubble expansion parameter is 
$H(z) = H_0 [\Omega_m (1+z)^3  + \Omega_{\Lambda}]^{1/2}$ in a flat Friedman cosmology 
($\Omega_m + \Omega_{\Lambda} = 1$).  All cosmological parameters and the resulting coefficients 
are based on WMAP-7 concordance values of CMB modeling from Komatsu \etal\ (2010), 
$\Omega_m h^2 = 0.1349 \pm 0.0036$, $\Omega_{\Lambda} = 0.728^{+0.015}_{-0.016}$, and  
$\Omega_b h^2 = 0.02260 \pm 0.00053$, and Hubble constant 
$H_0 = (100~{\rm km~s}^{-1}~{\rm Mpc}^{-1}) h$ with $h = 0.70$.  To set the ion densities $n_i$, we 
define $x_{\rm HI}$ and $x_{\rm HeII}$ as the abundance fractions of  \HI\ and \HeII, respectively.  
(The quantity $x_{\rm HI}$ is often called the neutral fraction.)  
The factors $\delta_i$ are the overdensities of H (at $z \approx 6$) and He (at $z \approx 3$) relative 
to the mean co-moving baryon density, 
$\langle \rho_b \rangle = (\Omega_b \rho_{\rm cr}) = 4.25 \times 10^{-31}$ g~cm$^{-3}$.
With a primordial helium abundance $Y = 0.2477 \pm 0.0029$ by mass (Peimbert \etal\ 2007) 
and $y = (Y/4)/(1-Y) = 0.0823$ by number, this baryon density corresponds to \HI\ and \HeII\ number 
densities,
\begin{eqnarray}
   n_{\rm HI}     &=& (6.55 \times 10^{-5}~{\rm cm}^{-3}) \;  x_{\rm HI} \;   \delta_{\rm H} 
       \left[  \frac {(1+z)}{7} \right] ^3  \; ,  \\
   n_{\rm HeII} &=& (1.01 \times 10^{-6}~{\rm cm}^{-3}) \;  x_{\rm HeII} \;  \delta_{\rm He}
       \left[  \frac {(1+z)}{4} \right] ^3    \;  . 
 \end{eqnarray}
In the  high-$z$ limit, $\Omega_m (1+z)^3 \gg \Omega_{\Lambda}$, we can express the Hubble
parameter as $H(z) \approx H_0 \Omega_m^{1/2} (1+z)^{3/2} \approx 
(294~{\rm km~s}^{-1}~{\rm Mpc}^{-1}) [(1+z)/4)]^{3/2}$ scaled to the epoch $z \approx 3$ relevant for 
the beginning of \HeII\ reionization.  

Thus, the Gunn-Peterson optical depths for H I (scaled to the $z = 6$ epoch) and He II (scaled to
the $z = 3$ epoch) are:
\begin{eqnarray}
      \tau_{\rm HI} (z)    &=& (3.99 \times 10^5) \left[ \frac {(1+z)}{7} \right] ^{3/2} \; x_{\rm HI} \;  \delta_{\rm H}       \; ,  \\ 
      \tau_{\rm HeII} (z) &=& (3.55 \times 10^3) \left[ \frac {(1+z)}{4} \right] ^{3/2} \; x_{\rm HeII} \;  \delta_{\rm HeII} \; .
\end{eqnarray}
Very small abundance fractions, $x_{\rm HI} \approx 10^{-4}$ and $x_{\rm HeII} \approx 10^{-2}$,
will saturate the Gunn-Peterson \Lya\ absorption in \HI\ and \HeII, with optical depths 
$\tau_i \gg 10$, even in under-dense regions with $\delta_i = 0.1$.  
It is primarily for this reason that the hydrogen reionization epoch is so difficult to measure 
in  \Lya\ absorption at $z \geq 6$ (Fan \etal\ 2006).  However, \HeII\ absorption at $z \leq 3$ can probe 
50--100 times higher abundance fractions in $x_{\rm HeII}$.  For the maximum observable values of 
\HeII\  optical depth ($\tau_{\rm HeII} = 5$) and with typical densities in low-density 
filaments and voids, the intervals of high \HeII\  absorption at $z = 2.8 \pm 0.1$ correspond
to \HeII\  fractions,
\begin{equation}
   x_{\rm HeII} = (0.015) \left[ \frac {\tau_{\rm HeII}} {5} \right] \left[ \frac {(1+z)} {3.8} \right] ^{-3/2}
      \left[ \frac {\delta_{\rm He}} {0.1} \right] ^{-1}   \; . 
\end{equation}
These overdensities, $\delta_{\rm He} = 0.1$,  are appropriate for the low-column-density 
($\log N_{\rm HI} \leq 13.5$) regions of the \Lya\ forest.  

 Figure 5 shows the \HeII\ optical depths for $2.4 < z < 2.9$,  using COS data from both
 G140L and G130M and \FUSE.   To construct this
 figure, we reconstructed ``horizontal error bars" representing the redshift bins ($\Delta z$)
 used in these papers.  We used  unpublished $\Delta z$ and uncertainties in optical depths 
 $\tau_{\rm HeII}$ from Shull \etal\ (2004).   From Zheng \etal\ (2004), we used values of 
 $\Delta z$ and $\tau_{\rm HeII}$ from their Figure~3, with 25\% uncertainties in $\tau_{\rm HeII}$.
 We find reasonable agreement with the \FUSE\ optical depths (Shull \etal\ 2004) down to $z = 2.4$, 
 although some of the G140L fluxes are higher than those in the \FUSE\ extension to $z \leq 2.4$ 
 (Zheng \etal\ 2004).   The smaller \FUSE\ optical depths may arise from uncertain background 
 subtractions.  Defining the redshift ranges for these data points is difficult for optical 
 depths that vary rapidly over $\Delta z \approx 0.01$.


\begin{figure*}
  \epsscale{1.2}
   \plotone{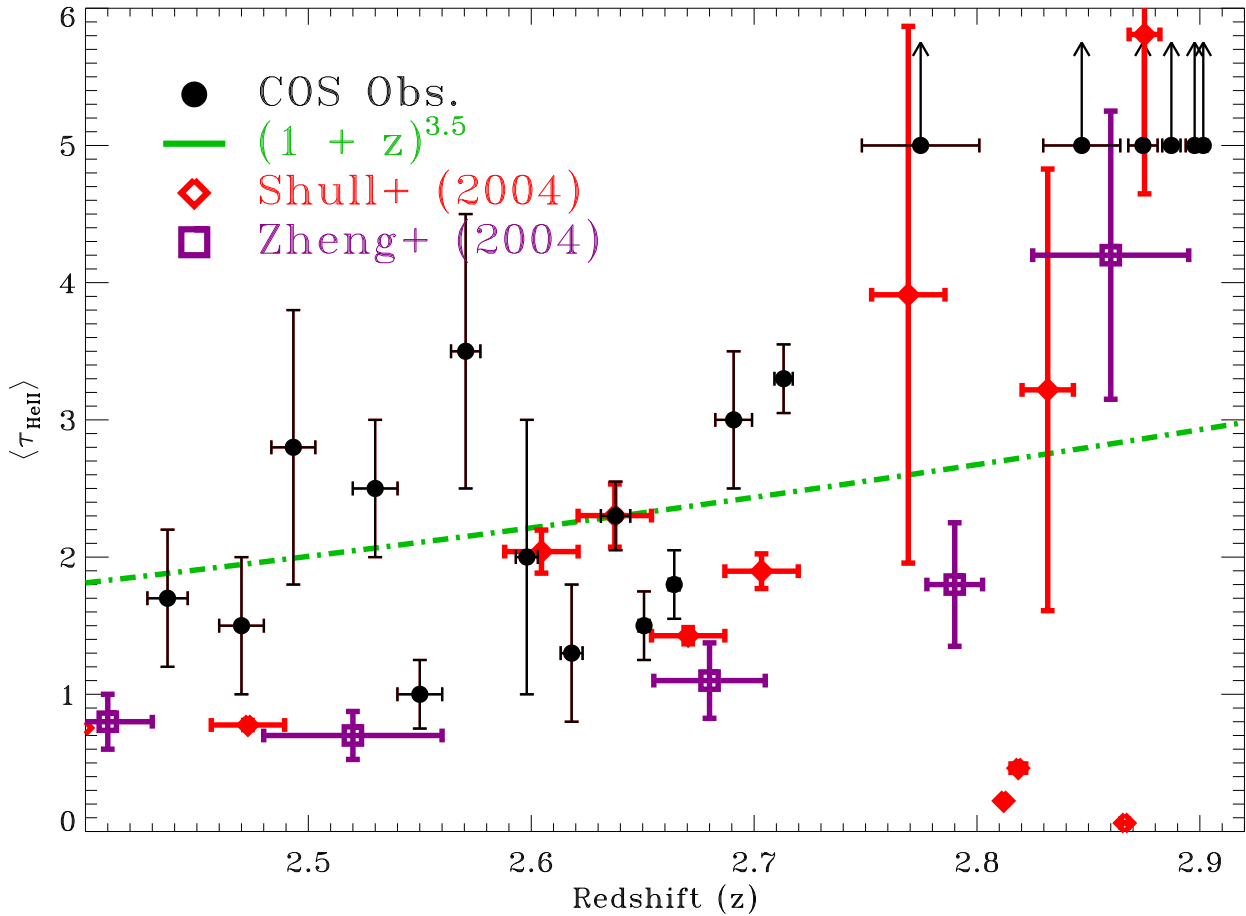}    
   \caption{Absorption optical depths, $\tau_{\rm HeII}$, vs.\ redshift (bins listed in Table 2) for
   this paper (black points: COS G130M and G140L) and from previous \FUSE\ measurements
   (Zheng \etal\ 2004; Shull \etal\ 2004).   COS optical depths are generally higher than those 
   found by \FUSE.  
   Theoretically predicted $(1+z)^{3.5}$ redshift scaling  (Fardal \etal\ 1998) is shown as a dotted line.  
   High COS optical depths, $\tau_{\rm HeII} \geq 5$, are plotted as lower limits,  based on zero-level 
   flux accuracy implied by RMS  variations, 
   $\sigma_F = 2.7 \times 10^{-17}$  ergs~cm$^{-2}$~s$^{-1}$~\AA$^{-1}$, at  the bottom 
   of troughs at 1145--1155 \AA\ and 1162--1173~\AA. }
\end{figure*}


 Future COS spectra of fainter \HeII-GP targets (often $19^{\rm th}$ magnitude AGN) will have low 
 S/N ratios and will probably not show the fluctuations in optical depths found here.   At best, they 
 will measure broad-band optical depths,  $\langle \tau_{\rm HeII} \rangle$.   For HE~2347$-$4342, 
 we list below the \HeII\ and \HI\ optical depths averaged over $\Delta z = 0.1$ windows, together 
 with the standard deviations of their distributions:  
\begin{itemize}
  
\item ($z =$ 2.4--2.5)    $\langle \tau_{\rm HI} \rangle  = 0.12^{+0.21}_{-0.09}$ ,
         $\langle \tau_{\rm HeII} \rangle = 1.11^{+0.97}_{-0.49}$  
        
\item ($z =$ 2.5--2.6)    $\langle \tau_{\rm HI} \rangle = 0.12^{+0.25}_{-0.08}$ ,
       $\langle \tau_{\rm HeII} \rangle = 1.18^{+0.71}_{-0.50}$ 

\item ($z =$ 2.6--2.7)  $\langle \tau_{\rm HI} \rangle = 0.26^{+0.36}_{-0.18}$ ,
     $\langle \tau_{\rm HeII} \rangle = 1.37^{+1.03}_{-0.53}$ 
    
\end{itemize} 
There may be a slight trend to higher $\eta$ at lower redshifts, as we discuss in Section~3.4.
Considering the ratios of the broad-band averages, we find that $\langle \eta \rangle$ rises from
$21^{+33}_{-17}$ at $z = 2.6-2.7$ to values $39^{+84}_{-31}$ at $z = 2.5-2.6$ and 
$37^{+72}_{-34}$ at $z = 2.4-2.5$.  Despite the large spreads of the optical-depth distributions,
the means are fairly well determined.  
 
Figure 6 shows examples of the \HeII\ and \HI\ ionization structures found in our cosmological simulations 
(Smith \etal\ 2010), computed with \texttt{Enzo}, an N-body,  hydrodynamical code modified to
compute time-dependent H and He ionization.  One can see the projected topology of the cosmic web, 
the baryon overdensity, the ionization fractions of \HI\ and \HeII\ at  $z = 2-3$, and their ratio,  
$x_{\rm HeII}/x_{\rm HI}$.   Observationally, we measure the ratio of column densities, 
$\eta =$ N(\HeII)/N(\HI), which we approximate as $\eta \approx 4 \tau_{\rm HeII} / \tau_{\rm HI}$.
The factor of 4 arises from the larger \HeII\ frequency bandwidth, with the absorption-line cross
section, $\sigma_{\lambda} \propto f \lambda$, scaling as oscillator strength times wavelength.  
Our IGM simulations at $z \approx 2.5$ show a mean value of the ionization ratio, 
$x_{\rm HeII}/x_{\rm HI} \approx 500$, corresponding to $\langle \eta \rangle \approx 40$ 
for primordial He/H = 0.0823 by number.  We have confirmed that a substantial amount of
observable \HeII, with $x_{\rm HeII} =$ 0.001--0.01, comes from ionized gas with 
$\log T = 4.0-5.3$ (much of it at $\log T = 4.3$) and  $\log \delta_H = -1$ to $+2$.  
This \HeII\ gas is primarily confined to photoionized filaments in the cosmic web, with some 
contribution from collisional ionization (at $T > 10^5$~K).   Further analysis of these simulations 
will appear in a later paper.


\begin{figure*}
   \epsscale{1.2}
   \plotone{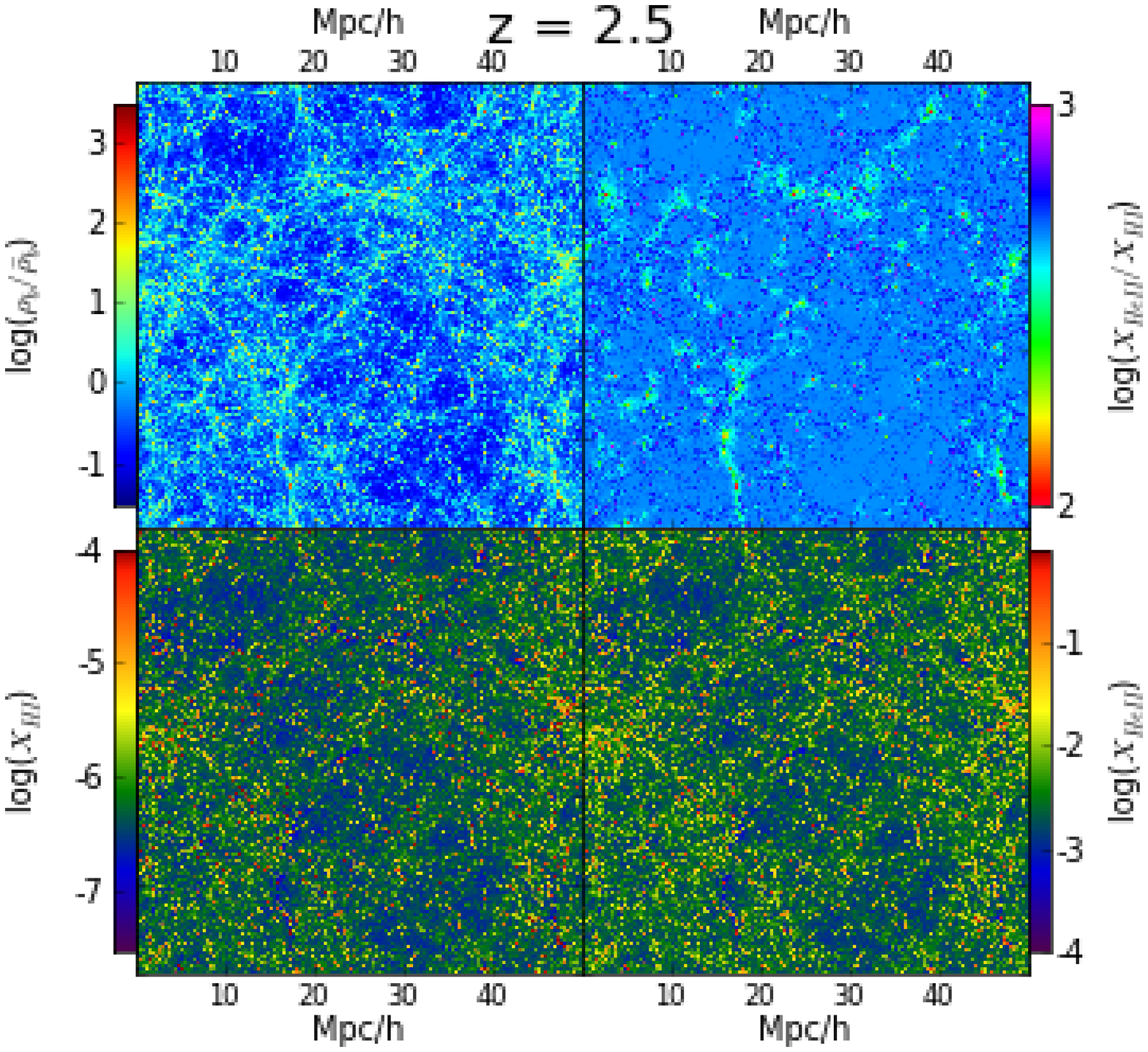}    
   \caption{Four-panel plot of results from our IGM simulations  at $z = 2.5$ (Smith \etal\ 2010) 
   showing baryon overdensity (top left), ionization fractions of H\,I and He\,II (bottom left and right), and
   their ratio $x_{\rm HeII}/x_{\rm HI}$ (top right).  Filamentary structures in the cosmic web are
   obvious in both top panels, with a mean value, $\langle x_{\rm HeII}/x_{\rm HI} \rangle \approx 500$,
   corresponding to $\eta = n_{\rm HeII}/n_{\rm HI} \approx 40$ for primordial He/H = $y$ = 0.0823.
   Much of the He\,II in these filaments appears to be photoionized gas at $10^{4.3}$~K, but with
   some contribution from collisional ionization at $T > 10^5$~K.   {\bf Original paper with higher
   resolution available at}   {\tt http://casa.colorado.edu/$\sim$mshull/HeII-paper.pdf}    }
\end{figure*}


\subsection{He~II and H~I Optical-Depth Ratios  }

The \HI\ and \HeII\ \Lya-forest absorbers have traditionally been assumed to arise in
photoionized gas, irradiated by metagalactic ionizing (EUV) radiation of specific intensity
$J_{\rm \nu} = J_i (\nu / \nu_i)^{-\alpha_s}$,  where $i$ refers to \HI\  ($h \nu_i = 13.6$ eV) 
or \HeII\ ($h \nu_i = 54.4$ eV).  For these hydrogenic species, whose photoionization cross 
sections above threshold $h \nu_i$ scale as $\sigma_{\nu}  = \sigma_i (\nu / \nu_i)^{-3}$, the 
photoionization rates $\Gamma_i  = (4 \pi \sigma_i/h) J_i / (\alpha_i + 3)$. The parameters 
$\alpha_i$ are local spectral indices of the ionizing background at 1 ryd and 4 ryd, respectively, 
and provide minor corrections to the photoionization rates. For typical metagalactic fluxes and 
IGM densities during reionization, hydrogen and helium are mostly ionized, with fractions 
$x_{\rm HI} \ll x_{\rm HII}$ and $x_{\rm HeI} \ll x_{\rm HeII} \ll x_{\rm HeIII}$.
Because the first ionization potentials of \HI\ (13.6~eV) and \HeI\ (24.58~eV) are not greatly 
disparate, we expect the \HeIII\ and \HII\ ionization fronts to coincide for appropriate 
values of ionizing spectral index.  For QSO ionizing spectra of power-law form, the \HeII\  and 
\HI\ ionization fronts, driven by the photon fluxes at 1 ryd and 4 ryd, respectively, will propagate 
at the same (photon flux-limited) velocity for a critical spectral index,
 \begin{equation}
       \alpha_{\rm crit} = - \frac {(\ln y) } {\ln 4} \approx 1.80  \;  ,
 \end{equation}
 corresponding to our adopted He/H abundance, $y = 0.0823$.
 The agreement with the \HST-observed mean spectral index of AGN at 1--2 ryd,
 $\langle \alpha_s \rangle = 1.76 \pm 0.12$ (Telfer \etal\ 2002), suggests that  \HII\ and
 \HeIII\ ionization fronts normally propagate together.    
  
 In photoionization equilibrium in regions of high ionization, $x_{\rm HI} \ll 1$ and $x_{\rm HeII} \ll 1$,  
 the \HeII/\HI\ abundance ratio is (Fardal \etal\ 1998;  Shull \etal\ 2004),
\begin{eqnarray}
   \eta  & \equiv & \frac {N_{\rm HeII}} {N_{\rm HI}} \approx 
         \frac {n_{\rm HeIII}} {n_{\rm HII}} \frac {\alpha^{(A)}_{\rm HeII}} {\alpha^{(A)}_{\rm HI}} \; 
         \frac {\Gamma_{\rm HI}} {\Gamma_{\rm HeII}}   \nonumber  \\
        & \approx & (1.77) \frac {J_{\rm HI}} {J_{\rm HeII}} 
         \frac {(3 + \alpha_{\rm HeII})} {(3+\alpha_{\rm HI})} \;  T_{4.3}^{0.042}   \; .  
\end{eqnarray}
The numerical coefficient has been increased from 1.70 to 1.77, reflecting an updated value
of the primordial He/H abundance, $y = 0.0823$, which we use as an approximation to 
$n_{\rm HeIII} / n_{\rm HII}$.   
In this equation, $\alpha^{(A)}_{\rm HI}$, $\alpha^{(A)}_{\rm HeII}$, $\Gamma_{\rm HI}$, and 
$\Gamma_{\rm HeII}$ are the case-A recombination rate coefficients and photoionization rates 
for \HI\ and \HeII, and $J_{\rm HI}$ and $J_{\rm HeII}$ are the specific intensities of the ionizing 
radiation field at 1 ryd and 4 ryd.  The parameter $T_{4.3} = (T / 10^{4.3}~{\rm K})$ 
is used to scale the temperature dependence of the Case-A recombination rates over the range
$T = 10,000$~K to 20,000~K (Osterbrock \& Ferland 2006):
$\alpha^{(A)}_{\rm HeII} \approx (1.36 \times 10^{-12}~{\rm cm}^3~{\rm s}^{-1})T_{4.3}^{-0.694}$ and
$\alpha^{(A)}_{\rm HI} \approx (2.51 \times 10^{-13}~{\rm cm}^3~{\rm s}^{-1})T_{4.3}^{-0.736}$.  

Equation (8) is an approximation, which will break down when $x_{\rm HeII}$ becomes significant. 
It also does not include the effects of collisional ionization of \HeII, which can become 
important in the shocked filaments of shocked gas at $T > 10^5$~K, the so-called warm-hot 
intergalactic medium (WHIM).   As noted in Figure 6, our simulations suggest that some of the
\HeII\ in shocked IGM filaments may be collisionally ionized gas at $T > 10^5$~K.  However,
the bulk of the \HeII\ is photoionized gas at $T \approx 10^{4.3 \pm 0.3}$~K.


\begin{deluxetable*}{ccccc}
\tabletypesize{\footnotesize}
\tablecaption{Statistics of  He\,II/H\,I Column-Density Ratios \tablenotemark{a} }
\tablewidth{0pt}
\tablehead{
 \colhead{Statistic} &
 \colhead{G140L} &
 \colhead{G130M} &
 \colhead{G140L} &
 \colhead{G130M}   \\
 \colhead{ } &
 \colhead{($2.4 < z < 2.73$)} &
 \colhead{($z>2.73$)} &
 \colhead{($2.73 < z <2.87$)} &
 \colhead{($2.73  <  z  < 2.87$)}  \\
 \colhead{ } &
 \colhead{Low-Res'n} &
 \colhead{Medium Res'n} &
 \colhead{Overlap Region} &
 \colhead{Overlap Region}   }
\startdata	
Mean ($\eta$)        &  $47 \pm42$            &   $209\pm281$   & $105\pm121$   & $208\pm281$      \\
Median ($\eta$)      &  $33^{+56}_{-19}$      &$88^{+391}_{-71}$& $69^{+94}_{-41}$&$87^{+388}_{-71}$ \\
Mean ($\log \eta$)   &  $1.52\pm0.38$         &  $1.90\pm0.68$  & $1.83\pm0.41$   & $1.89\pm0.68$    \\
Median ($\log \eta$) &  $1.51^{+0.43}_{-0.39}$&  $1.94\pm0.74$  & $1.84\pm0.38$   & $1.94\pm0.74$    \\
\enddata

\tablenotetext{a}{We list mean and median values of the ratio $\eta
\equiv N_{\rm HeII} /N_{\rm HI}$ over redshift ranges covered at low
resolution (G140L) and medium resolution (G130M).  The last two
columns show $\eta$ statistics for the overlap range, $2.73 < z <
2.87$.  All statistics are calculated including both saturated and
unsaturated values of $\tau_{\rm HeII}$.   Using only unsaturated pixels
reduces $\langle \eta \rangle$ by 10--20\% at $z > 2.7$, with no change
at $ z < 2.7$.  }

\end{deluxetable*}



\subsection{Troughs of Strong \HeII\ Absorption} 

As shown in Figures 2 and 4, we have confirmed the presence of three long troughs
with strong \HeII\ absorption ($\tau_{\rm HeII} \ga 5$) and no detectable flux.  These troughs 
have been noted previously (Reimers \etal\ 1997; Smette \etal\ 2002; Shull \etal\ 2004),
but the very low COS backgrounds allow us to better characterize their optical depth.  We also
detect numerous  windows of flux transmission (e.g., at 1138, 1139, 1144, 1154,
1160, 1174, and 1179 \AA\ -- see Figure~2).  
The G130M data provide accurate boundaries in redshift.  From those, we can determine 
their comoving radial sizes, using WMAP-7 cosmological parameters:
\begin{enumerate}

\item {\it (Trough 1)}  $z = 2.751-2.807$ (61 Mpc), 
  
\item  {\it (Trough 2)} $z = 2.823-2.860$ (39 Mpc),
  
\item  {\it (Trough 3)} $z = 2.868-2.892$ (25 Mpc). 

\end{enumerate}  
These dimensions, all much larger than 10 Mpc, are incompatible with zones of photoionization 
around typical QSOs. They are also unlikely to exist at the epoch following the overlap of expanding 
cosmological I-fronts.   Recent theoretical modeling has studied the statistics of \HeII\ troughs 
(McQuinn \etal\ 2009; McQuinn 2009;  Furlanetto 2009a,b; Dixon \& Furlanetto 2009).  These
studies examine large-scale fluctuations in the \HeII\ \Lya\ forest transmission, during and after
\HeII\ reionization.   A robust result is that long troughs of large optical depth, $\tau_{\rm HeII} > 4$, 
cannot be explained in models with a smoothly evolving ionizing radiation field (Furlanetto \& Dixon 2010).  
 Furthermore, the observed 25--60 Mpc troughs of high \HeII\ opacity between $2.751 < z < 2.892$ cannot 
 be accommodated in standard reionization models unless \HeII\ reionization is delayed until $z < 2.7$.  
 The \HeII\ absorption is therefore expected to be patchy at $z \ga 2.7$, as we have observed. 

It would be useful to conduct surveys (Worseck \etal\ 2007) of galaxies and QSOs 
near the sight line at these trough redshifts to identify potential sources of ionization (or lack 
thereof).    As discussed in Section 3.2, the \HeII\ opacity greatly exceeds that of \HI;  the ratio 
$\eta \gg 1$ for typical conditions in an IGM photoionized by QSOs of low space density.  Thus, 
the propagation and overlap of I-fronts driven by \HeII\ ionizing continua will likely differ from those
of hydrogen.  The greater \HeII\ opacity means that filaments of the cosmic web still have 
$\tau_{\rm HeII} = 2-3$ at $z = 2.4-2.7$ (see Figures 5 and 6).   
Thus, the topology of overlapping  \HeIII\ ionized zones probably differs from that of merging
\HII\ regions at $z = 6$.  These theoretical results are explored in a later paper (Smith \etal\ 2010) 
using adaptive-mesh cosmological simulations with the \texttt{Enzo} code.


\subsection{Fluctuations in He~II/H~I Ratios.}  

Because reionization involves the stochastic propagation and overlap of ionization fronts
(I-fronts), one naturally expects variations in the abundances of \HI\ and \HeII.   The metagalactic 
flux of high-energy (4 ryd continuum) radiation varies primarily because of the low space density 
of the dominant ionizing sources (AGN) and varying attenuation through the IGM.  The variations 
in \HeII\ opacity can be especially large during reionization, because the 4 ryd continuum is strong 
near AGN (inside the I-fronts) but heavily attenuated in the \HeII\ troughs, far from the AGN and 
within the gaseous filaments of the cosmic web.  

Figure 4 shows both \HeII\ and \HI\ absorption, overlaid in redshift,  illustrating the correspondence 
of \HeII\ flux-transmission windows with weak spots (low \HI\ absorption) in the \HI\ \Lya\ forest.   
The non-Gaussian wings in the COS line-spread function (LSF) is most significant for weak 
absorbers, but we believe this is completely accounted for in the convolution of the \HI\ data by 
the COS LSF.  The resolving power ($R = 18,000$) of the COS LSF is more appropriately the 
effective Gaussian resolving power (Green \etal\ 2010).   The main difference in the convolved 
and unconvolved \HI\ data in Figure 4 is simply a degradation in resolution by a factor of 2.5.  
The COS LSF contributes to a redistribution of flux in the narrowest lines, but this is 
accounted for with the non-Gaussian LSF.

The \HeII/\HI\ ratio, $\eta \approx 4 \tau_{\rm HeII} / \tau_{\rm HI}$, tracks the ratio of ionizing fluxes 
at 1 and 4 ryd (see equation [8]).  Table 3 presents the mean and median values of $\eta$ in the
G130M and G140L data, and over various redshifts.  In different columns, we present these statistics 
for the G140L and G130M spectral regions, and for the region ($2.73 < z < 2.87$) where the two 
gratings overlap.  To calculate $\rm \eta(z)$, we convolve the high-resolution VLT data with either
the medium-resolution (G130M) or low-resolution (G140L) COS line-spread functions, depending 
which dataset is representing \HeII\ at a particular redshift.  The COS data are then smoothed by 
seven pixels ($\sim1$ resolution element) and up-sampled to the finer VLT wavelength scale.
Optical depths $\tau_{\rm HeII}(z)$ and $\tau_{\rm HI}(z)$ are calculated on a pixel-by-pixel basis.


\begin{figure*}
  \epsscale{1.2}
   \plotone{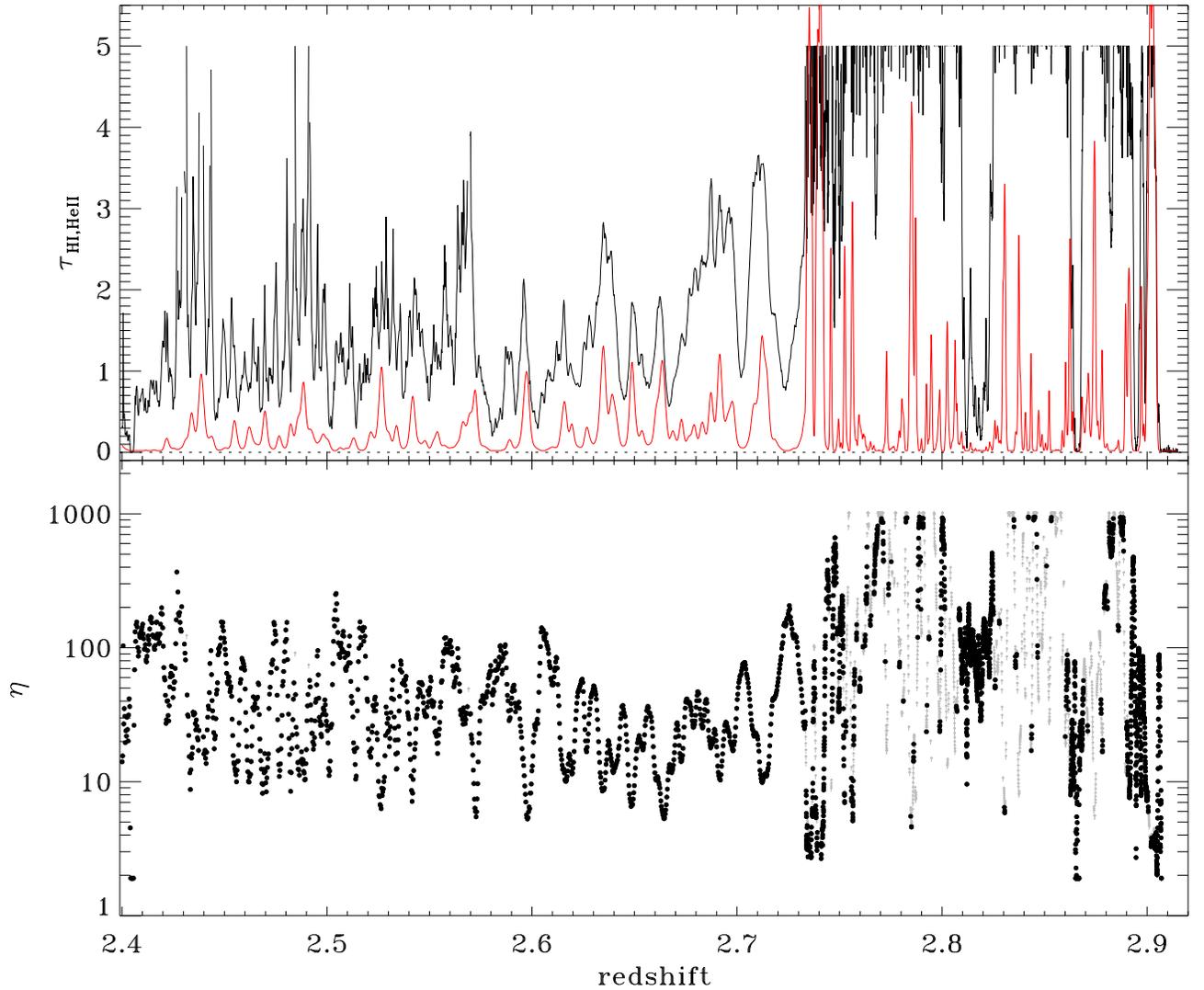}    
   \caption{Optical depth for He\,II (black) and H\,I  (red) and their ratio,
   $\eta \approx 4 \tau_{\rm HeII} / \tau_{\rm HI}$, versus redshift.  Note the change in 
   spectral resolution at $z < 2.73$, the transition between G130M and G140L gratings.  
   The VLT/UVES data have been convolved with the appropriate COS line spread function.  
   We only plot values $\tau_{\rm HeII} \leq 5$.  Fluctuations at $\Delta z \approx 0.01$ 
   correspond to proper distances of 10.8~Mpc at $z = 2.8$.   For our detectable ranges of 
   optical depth ($0.02 \leq \tau_{\rm HeII} < 5$ and $0.01 < \tau_{\rm HI} < 5.5$) we
   believe $\eta$ can be determined reliably over the range $5 < \eta < 500$ (G130M)
   and $10 < \eta <100$ (G140L).  
   The mean values and standard deviations are $\langle \eta \rangle = 47 \pm 42$ (for G140L 
   range) and $\langle \eta \rangle = 189 \pm 276$ (for G130M range).  See full statistics in 
   Table 3.  There may be a slight trend of higher $\eta$ toward lower $z$. }
\end{figure*} 



\begin{figure*}
  \epsscale{1.2}
   \plotone{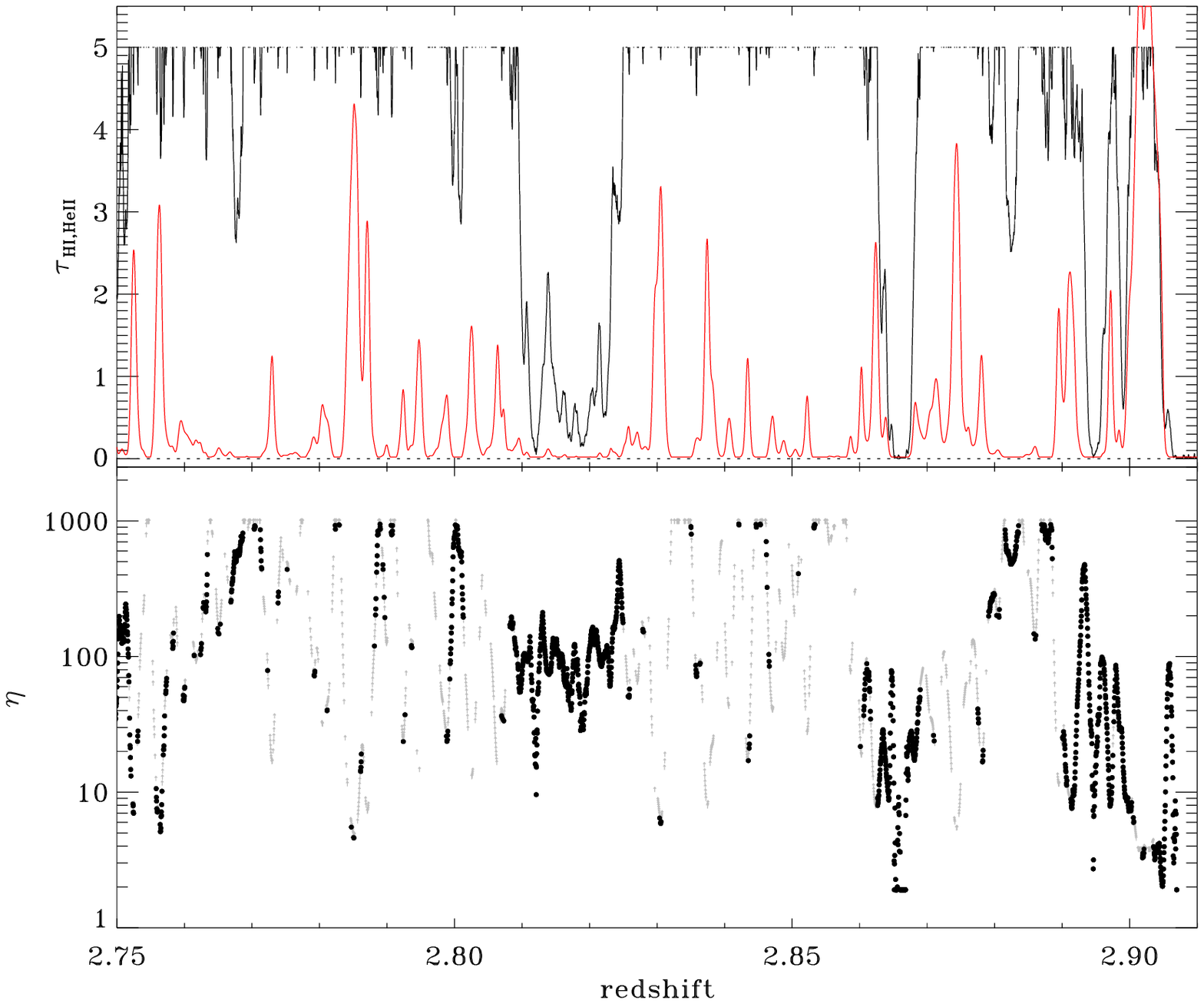}    
   \caption{Closeup of optical depth for He\,II (black) and H\,I (red) and their ratio $\eta$ at redshifts 
   near the QSO, $2.75 \leq z  \leq 2.90$.  In the higher-resolution G130M data, 
   $\eta$ varies on scales down to $\Delta z \approx 0.003$.  Grey points indicate locations where
   $\tau_{\rm HeII} \geq 5$, so that $\eta$ is a lower limit.  We believe $\eta$ can reliably be measured
   over the range $5 < \eta < 500$, considering reasonable limits on $0.02 < \tau_{\rm HeII} < 5$ and
   $0.01 < \tau_{\rm HI} < 5.5$.  }
\end{figure*}


Figures 7 and 8 show $\tau(z)$ and $\eta(z)$ and demonstrate that  \HeII\  optical depth saturates 
in certain regions.  Over the range $2.4 < z < 2.9$, 32\% of pixels are saturated, but this 
fraction depends strongly on redshift and spectral resolution.   In the low-resolution data at 
$z < 2.73$, we find that $\tau_{\rm HeII}$ is almost always unsaturated.  At  $z > 2.7$, 
we see long absorption troughs in which 40--45\% of the pixels have $\tau_{\rm HeII} >5$.
We computed mean and median values of $\eta$ using both saturated and unsaturated pixels.
As the saturated pixel values for \HeII\ are entirely in the numerator, our average $\eta$ values 
are lower limits.  Using only unsaturated pixels reduces $\langle \eta \rangle$ systematically by 
10-20\% at $z>2.7$.  Correcting for larger true values of $\tau_{\rm HeII} > 5$ would increase
the mean statistics.  There is no change at $z<2.7$.

We confirm the fluctuations seen by \FUSE\  between $2.73 < z <2.90$ (Kriss \etal\ 2001; 
Shull \etal\ 2004) on small scales, $\Delta z \approx 0.003-0.01$.  Figures 7 and 8 show  
that $\eta$ fluctuates on scales down to $\Delta z \approx 0.01$ at G140L resolution and
$\Delta z \approx 0.003$ at G130M resolution.  For reference, the redshift interval $\Delta z = 0.01$ 
corresponds to a comoving radial distance of 10.8 Mpc at $z = 2.8$.   
Regions of hot gas ($T > 10^5$~K) might appear as high-$\eta$ systems, in which 
\HI\ is far more collisionally ionized than \HeII.  However, the COS data also exhibit 
several instances of low-$\eta$ absorbers with $\eta \approx 5-20$.  
Since $\eta \approx 4 \tau_{\rm HeII} / \tau_{\rm HI}$, the low-$\eta$ systems appear primarily where
the \HI\ optical depth is large and \HeII\ optical depth small.  If photoionized, these absorbers may be
irradiated by very hard radiation fields, with large ratios of $J_{\rm HeII} / J_{\rm HI}$ (equation 8), 
perhaps near an AGN with a particularly flat EUV spectral index,  $\alpha_s < 1$.  

Several good examples of low--$\eta$ absorbers appear at $z \approx 2.755$, 2.785, 2.830, 2.865, 
and 2.904.   Most of these systems have high $\tau_{\rm HI}$, the exception being the absorber at 
$z = 2.865$ in which both $\tau_{\rm HeII}$ and $\tau_{\rm HI}$ are low, and their ratio poorly determined.  
For high-$\eta$ systems, the G130M data exhibit regions with $\eta = 200-500$, typically in gas
with large $\tau_{\rm HeII}$ but low $\tau_{\rm HI}$.    Extreme excursions in $\eta$ may arise from 
noise, and their detectable range depends on spectral resolution.  We therefore take care to
assess the range over which we can measure reliable optical depths.  For ratios within the range 
$\eta =$ 10--100 (G140L) and $\eta =$ 5--500 (G130M), we believe nearly all the fluctuations
are real.  Unfortunately, strong troughs of \HeII\ absorption at $z > 2.74$ hinder our ability
to measure $\tau_{\rm HeII}$ for individual systems with the high-resolution G130M.
Many of the $\eta$-fluctuations appear at $z < 2.7$, observed with the lower-resolution G140L.  

These small-scale $\eta$-variations have been the topic of numerous theoretical papers
(Zuo 1992; Fardal \& Shull 1993; Fardal \etal\ 1998; Bolton \etal\ 2005, 2006).  The key focus
of these papers is on fluctuations in the small numbers of AGN within the characteristic 
attenuation length, corresponding to the $``\tau = 1$ sphere" at 4 ryd photon energy.  
Because the \HeII\ opacity is so much larger than that of \HI, one expects large variations 
in the \HeII\ ionizing flux, $J_{\rm HeII}$, on comoving scales of 20--40 Mpc at
$z \approx 2.4-2.9$ (Fardal \etal\ 1998; Furlanetto 2009; Dixon \& Furlanetto 2009).  
A second important effect is that the observed low-redshift AGN exhibit wide variations in their 
EUV spectral indices (Telfer \etal\ 2002; Scott \etal\ 2004), which produce strong source 
variations in the flux ratio $J_{\rm HeII}/J_{\rm HI}$.  
Finally, radiative transfer of the metagalactic ionizing radiation reprocesses the ionizing 
spectra, producing further variations (Fardal \etal\ 1998).  What is now firmly established  by 
the new COS data are the variations in $\eta$ on $\sim10$ Mpc comoving radial distance scales 
($\Delta z \approx 0.01$).


\subsection{Proximity Effect and Quasar Str\"omgren Spheres}    

It is surprising to note the sharpness of the \HeII\ edge at 1186.26 \AA\  and the lack of
any proximity effect  near the QSO.   If ionizing radiation were escaping from the AGN,
the surrounding IGM should be ionized within 5--10 Mpc of the QSO.  
The \HeII\ absorption extends $\Delta z \approx 0.019$ beyond the previous QSO redshift 
estimate, $z_{\rm em} = 2.885$ (Reimers \etal\ 1997).   As discussed in Section 3.1, this 
dilemma is easily resolved by moving the quasar redshift to $z = 2.904$, an offset 
(1460 \kms\ at $z = 2.9$) that corresponds to 19.5 Mpc comoving proper distance.   
However, the two proposed systemic redshifts, $z_{\rm QSO} = 2.885$ and 2.904, 
raise several questions.   Does the QSO  lie inside an \HII\ and  \HeIII\ cavity produced by 
its photoionizing radiation?  Or, is the QSO ionizing radiation absorbed by circumnuclear gas?  
Is there infalling gas near the AGN that produces associated absorption with  
$z_{\rm abs} > z_{\rm em}$?  It is worth investigating these possibilities.  

For \HI\ (1 ryd) and  \HeII\ (4-ryd) ionizing-photon luminosities, $S_0^{\rm (H)}$ and
$S_0^{\rm (He)}$, what Str\"omgren radii will this bright QSO carve out?  These
distances are obviously limiting cases, in ionization equilibrium, whereas one expects
the ionization zones around the QSO to involve time-dependent propagation of
ionization fronts.  Indeed, the lack of a proximity effect could arise from this source
just starting to burn its way out of the galactic nucleus.  
Given the strength of the $z = 2.9$ absorber, we will assume higher gas densities 
associated with the QSO and scale to an overdensity ($\delta_H \approx 100$) at $z = 2.9$.   
The H and He densities are then: 
$n_{\rm H} = (1.13 \times 10^{-3}~{\rm cm}^{-3}) (\delta_{\rm H}/100)$, $n_e \approx 1.165 n_H$, 
and $n_{\rm He} = (9.30 \times 10^{-5}~{\rm cm}^{-3}) (\delta_{\rm H}/100)$.   At  $T = 10^{4.3}$~K, 
$\alpha_{\rm HeII}^{(B)} = 9.1 \times 10^{-13}~{\rm cm}^3~{\rm s}^{-1}$ and
$\alpha_{\rm HI}^{(B)} = 1.4 \times 10^{-13}~{\rm cm}^3~{\rm s}^{-1}$.
A typical $(L^*)$ quasar produces a total of $\sim 10^{56}$ LyC photons s$^{-1}$,
but this QSO is much brighter ($V = 16.1$).   Extrapolation from 1200~\AA\  down to the \HeII\ 
Lyman limit (889.5 \AA\ at $z_{\rm QSO} = 2.904$) shows that HE~2347-4342 has an observed 
specific flux, $F_{\lambda} \approx 6 \times 10^{-15}$ erg~cm$^{-2}$~s$^{-1}$~\AA$^{-1}$  or
$F_{\nu} \approx 1.6 \times 10^{-27}$ erg~cm$^{-2}$~s$^{-1}$~Hz$^{-1}$.  From this flux 
and a luminosity distance, $d_L \approx 7.7 \times 10^{28}$ cm at $z = 2.904$, we estimate an 
ionizing photon luminosity of 
$S_0^{\rm (He)} \approx \nu L_{\nu}/3 \approx 1.6 \times 10^{57}$ photons~s$^{-1}$
in the 4-ryd continuum.  
Scaling to $S_0^{\rm (He)} \approx 10^{57}~{\rm photons~s}^{-1}$, we estimate a \HeIII\ 
Str\"omgren radius,
\begin{eqnarray}
    R_{\rm Str}^{(\rm HeIII)} & = & \left[ \frac {3 S_0^{\rm (He)}} 
          {4 \pi \, n_e \,  n_{\rm He}\,  \alpha_{\rm HeII}^{(B)} } \right] ^{1/3}   \nonumber \\
          & \approx &  (4~{\rm Mpc})  \left[ \frac {S_0^{{\rm (He)}} }{10^{57}~{\rm s}^{-1}} \right]^{1/3} 
           (\delta_H/100)^{-2/3}  T_{4.3}^{0.255}   \; .    \nonumber
\end{eqnarray}
A similar calculation, using the observed flux,
$F_{\lambda} \approx 1.2 \times 10^{-15}$ erg~cm$^{-2}$~s$^{-1}$~\AA$^{-1}$, 
at the redshifted (3561 \AA)  \HI\  edge, gives 
$S_0^{\rm (H)} \approx \nu L_{\nu}/3 \approx 5 \times 10^{57}$ photons~s$^{-1}$ and 
an \HII\ Str\"omgren radius of 
\begin{eqnarray}
    R_{\rm Str}^{(\rm HII)} & = & \left[ \frac {3 S_0^{\rm (H)}} 
          {4 \pi \, n_e \,  n_{\rm H}\,  \alpha_{\rm HI}^{(B)} } \right] ^{1/3} \nonumber \\
        & \approx & (6~{\rm Mpc})  \left[ \frac {S_0^{{\rm (H)}} }{5\times10^{57}~{\rm s}^{-1}} \right]^{1/3}   
           (\delta_H/100)^{-2/3}  T_{4.3}^{0.286}  \; .  \nonumber
\end{eqnarray}

We note that the ratio of the extrapolated fluxes at the respective ionization limits is fairly small.
In the QSO rest-frame, $F(912~{\rm \AA})/F(228~{\rm \AA}) \approx 5$, indicating a fairly hard (flat)
power-law spectrum $F_{\nu} \propto \nu^{-1.2}$.  These estimated Str\"omgren radii are a 
substantial fraction of the proper distance of 19.5 Mpc  between 
the two proposed QSO redshifts, $z = 2.904$ and $z = 2.885$.   Even though the local ionization
zone may not extend as far as these equilibrium Str\"omgren radii, the expected proximity effect is 
minimally visible, except for a narrow feature with $\sim 60$\% flux transmission at $z = 2.9055$, 
just below the \HeII\ edge (Figure 4).  We conclude that ionizing radiation from this QSO is
strongly absorbed in the proximity of the nucleus, possibly by circumnuclear gas or by the strong 
absorber at $z \approx 2.9$.  
The photoionization cross sections for \HI\ and \HeII\ above threshold are approximately
$\sigma_{\rm HI} = (6.30 \times 10^{-18}~{\rm cm}^2)(E/13.6~{\rm eV})^{-3}$ and
$\sigma_{\rm HeII} = (1.58 \times 10^{-18}~{\rm cm}^2)(E/54.4~{\rm eV})^{-3}$.  Thus, 
to absorb a substantial amount of the ionizing continua requires column densities 
\begin{eqnarray}
    {\rm N}_{\rm HI}    &=& (1.59 \times 10^{17}~{\rm cm}^{-2}) \left[ E/13.6~{\rm eV}\right] ^3 \; \tau(E)   \nonumber \\
    {\rm N}_{\rm HeII} &=& (6.35 \times 10^{17}~{\rm cm}^{-2}) \left[ E/54.4~{\rm eV}\right] ^3 \; \tau(E) 
\end{eqnarray} 
for continuum optical depth  $\tau$ at photon energy $E$.  

For this very bright QSO,  the absence of a  \HeII\ proximity effect indicates a significant
attenuation of the 4 ryd continuum, with optical depth $\tau > 3$ at mean photon energies 
1.5 times the 54.4~eV threshold, or N$_{\rm HeII} >  6 \times 10^{18}~{\rm cm}^{-2}$.  
What are the observational consequences of such large column densities of ionized gas? 
It has been suggested (Reimers \etal\ 1997) that  \OIII\ resonance lines at 303.415~\AA\ and 
305.596~\AA\ might be detectable, but we find no obvious candidates.  
It would be useful to search for signs of splittings of these lines, which are separated by 
2.182~\AA$(1+z) \approx 8.5$ \AA\ at $z \approx 2.9$.  
Some of the associated absorbers have \HI\ anomalously stronger than \HeII, suggesting that 
the gas is close to the QSO with helium mostly fully ionized (\HeIII).  
More clues may come from the strong \HeII\ absorbers at $z = 2.900-2.904$ (1184.7--1186.2~\AA)
which would be closest to the QSO if $z_{\rm QSO} = 2.904$.  
This system has high optical depth in both \HeII\ and \HI\ (see Figure 4).  Very strong metal-line
absorption is seen in \OVI\, with weaker absorption  in \CIV, \CIII, and \NV\ 
(Reimers \etal\ 1997; Fechner \etal\ 2004).  By fitting higher Lyman-series absorbers,
Fechner \etal\ (2004) found log N$_{\rm HI} = 16.02 \pm 0.03$, while the metal-line absorbers 
gave log N$_{\rm CIV} = 13.66 \pm 0.03$, log N$_{\rm NV} = 13.42 \pm 0.04$, and 
log N$_{\rm OVI} \ga 14.62$.  They proposed that this absorber is exposed to the strongest and 
hardest radiation from the QSO, and therefore may shield the other associated absorbers.  However,
their photoionization models gave log N$_{\rm HeII} \approx 16.3$, which is 100 times too small
to provide the necessary absorption (eq.\  [9]).  The \HI\ column is 10 times too small
to shield the 1 ryd continuum.   

Fox, Bergeron, \& Petitjean (2008)  obtained newer (high S/N) spectra on these 
metal-line absorbers at $z = 2.8916$, 2.8972, and 2.9041.  They found somewhat higher
column densities of \CIV\ and \NV\ and  a slightly higher column density,
$\log N_{\rm HI} = 16.29 \pm 0.05$ in the $z = 2.9041$ absorber.  This column is still insufficient to
shield the QSO's ionizing continuum.  We conclude that no satisfactory explanation exists for the 
absence of a proximity effect around this very luminous QSO, other than the possibility that
it has only recently turned on (within the last Myr).

  
\section{Conclusions and Implications} 

The high throughput and low background of the Cosmic Origins Spectrograph allow us to probe the
 \HeII\ reionization epoch toward HE~2347$-$4342 at redshifts $z$ = 2.4--2.9.  
Because of photoionization and recombination rates, \HeII\ is a far more abundant species than
\HI, with typical ratios $\eta = n_{\rm HeII}/n_{\rm HI}$ varying from 10--200 in the filaments of the \Lya\
forest.  With COS, we find that reionization of \HeII\ to \HeIII\ is patchy at $z = 2.7-2.9$, with 
5--10~\AA\ troughs of high optical depth ($\tau_{\rm HeII} > 5$) punctuated by narrow windows of 
flux transmission.
Previous studies of this QSO sight line (Kriss \etal\ 2001; Shull \etal\ 2004; Zheng \etal\ 2004)
found a gradual decrease in \HeII\ optical depth at $z < 2.7$,  suggesting a \HeII\ reionization
epoch $z_r = 2.8 \pm 0.2$.  With our new COS data, we find a more complicated picture, in
which \HeII\ is slowly reionized at $z \leq 2.7$.  

\noindent
We now summarize the new observational results and astrophysical issues:
\begin{enumerate}

\item  We have found clear evidence that the epoch of \HeII\ reionization is delayed below $z = 3$, 
   as we see patchy ionization down to $z \approx 2.7$.  For this bright AGN, high-S/N, low-background 
   observations from COS allow us to measure \HeII\ absorption, the UV continuum of the QSO, and 
   troughs of high-$\tau_{\rm HeII}$.  The high spectral resolution of the G130M grating probes absorption 
   fluctuations and the \HeII\ ionization edge, and the G140L grating allows us to study \HeII\ 
   reionization down to $z \approx 2.4$. 

\item Comparison of COS/G130M and G140L data with previous \HST\ and \FUSE\ data confirms 
   the far-FUV continuum,  absorption troughs, transmission windows, and \HeII\ optical-depth 
  below the \HeII\ edge at 1186.26~\AA.   The COS/G140L fluxes agree fairly well with \FUSE\
  data at $z > 2.7$ (Shull \etal\ 2004), but below 1130~\AA\  ($z < 2.7$) considerable differences 
  appear.  The COS fluxes are lower than those of FUSE inferred by Zheng \etal\ (2004), and our 
  derived \HeII\ optical depths are therefore somewhat higher at $2.4 < z <2.7$.  The broad-band 
  optical depths (averaged over $\Delta z = 0.1$) only drop below 
  $\langle \tau_{\rm HeII} \rangle \leq 1$ at $z < 2.4$. 
   
\item We observe optical-depth variations in \HeII\ and \HI\ that imply fluctuations in the
   \HeII/\HI\  abundance ratio on scales $\Delta z \approx 0.003-0.01$, corresponding to 
  4--10 Mpc in comoving distance.  Such variations likely arise from AGN source variations
  (1--4 ryd spectral indices) and IGM radiative transfer effects.  The number of low-$\eta$ 
  systems (mostly low $\tau_{\rm HeII}$ and high $\tau_{\rm HI}$) seen with COS is lower
  than claimed in FUSE observations.

\item We see intervals of patchy \HeII\  flux-transmission, as well as three long (5--10 \AA) 
   intervals of strong \HeII\ absorption (25--60 Mpc proper distance) between $z = 2.751-2.807$, 
   $z = 2.823-2.860$, and $z = 2.868-2.892$.  These troughs probably arise from dense filaments 
   in the baryon distribution and the scarcity of strong photoionizing sources (AGN) within 30--50 Mpc.  

\item  At $z =$ 2.7--2.9, probes of \HeII\ fraction, 
   $x_{\rm HeII} \geq (10^{-2})(0.1/\delta_{\rm He})(\tau_{\rm HeII} /5)$ 
    are now possible at optical depths $\tau_{\rm HeII} \geq 5$.  The  \HeII\  absorption
    is 50--100 times more sensitive to trace abundances than \HI\ at $z = 6$, where a neutral fraction
    $x_{\rm HI} = 10^{-4}$ saturates the \Lya\ Gunn-Peterson absorption.

\end{enumerate}

What future work is needed to better understand the \HeII\ reionization?  
First, we need to assess whether the HE~2347$-$4342 sight line is typical.  The
COS-GTO team will be observing two more \HeII\ AGN targets, HS1700+6416 and Q0302-003,
which may clarify whether the patchy reionization at $z < 2.8$ is common to most sight lines.  
Second, along the current sight line, we have identified many regions of partial flux transmission 
and three long troughs of \HeII\  absorption, where surveys of galaxies and AGN will help
clarify the sources of ionization.  Third, infrared spectra of the [\OIII] $\lambda$5007, 4959 
narrow emission lines (at 1.955 $\mu$m and 1.936 $\mu$m) could confirm the QSO systemic redshift.  
Pinning down $z_{\rm QSO}$ is crucial for understanding the nature of the ionization state
of the gas near the QSO, which shows no evidence of a proximity effect.  
On the theoretical front, cosmological simulations with time-dependent H and He chemistry
and I-front radiative transfer should elucidate the structure topology of the filaments and voids
that we observe in the \HeII\ \Lya-forest and \HeII\ troughs.  

Serious studies of  \HeII\ reionization have been hindered by the lack of many AGN targets
sufficiently bright to obtain far-UV spectra.  The spectrographs on \FUSE\ and \HST\ have
made good progress on three AGN sight lines, but newly discovered \HeII\ GP targets 
(Syphers \etal\  2009a,b) are considerably fainter than HE~2347$-$4342 or HS1700$+$6416.
New observations
of these fainter targets will probably measure only broad-band \HeII\ optical depths, for which
the high-resolution COS spectra may provide guidance in interpretation.  The long-term
study of the \HeII\ reionization process, from $z = 2.3-3.3$, must await the construction of
a future large-aperture, far-UV telescope in space (Shull \etal\  1999;  Postman \etal\ 2009).  
It would be especially useful if this mission had high sensitivity and moderate spectral resolution
($R \ga 20,000$) extending from 1130 \AA\ down to the 912~\AA\  hydrogen Lyman Limit.
This would enable the study of patchy \HeII\ reionization in \HeII\ $\lambda 304$ down to redshifts 
$z \ga 2$.


\medskip


\acknowledgments

It is our pleasure to acknowledge the thousands of people who made HST Servicing Mission 4 
a huge success.  We thank Brian Keeney, Steve Penton, St\'ephane B\'eland, and the rest of the 
COS/GTO team for their work on the calibration and verification of the early COS data, and 
Cristina Oliviera for her helpful input on G140L segment-B CALCOS processing.   We thank Dieter
Reimers for discussions regarding the QSO systemic redshift and for providing VLT data on the 
\HI\ \Lya\ forest.  Mark Giroux,  Blair Savage, David Syphers, and the referee provided numerous 
comments that improved the arguments made in this paper.  
This work is based on observations made with the NASA/ESA {\it Hubble Space Telescope}, obtained 
from the data archive at the Space Telescope Science Institute. STScI is operated by the Association 
of Universities for Research in Astronomy, Inc. under NASA contract NAS5-26555.
Our work was also supported by NASA grants NNX08AC146 and NAS5-98043 
and the Astrophysical Theory Program (NNX07-AG77G  from NASA and
AST07-07474 from NSF) at the University of Colorado at Boulder.

\end{document}